\documentclass[journal=jctcce,manuscript=article]{achemso}
\usepackage{amsmath}
\usepackage{bm}
\usepackage{graphicx,color}
\usepackage{stix}
\usepackage{subfigure}
\usepackage[usenames,dvipsnames]{xcolor}
\usepackage{array}
\usepackage{makecell}
\usepackage{algorithm}
\usepackage{algpseudocode}

\newcommand*{\GW}{{\textit{GW}}}
\newcommand*{\gw}{{\textit{G}\textsubscript{0}\textit{W}\textsubscript{0}}}

\newcommand{\eri}[2]{{\left( #1 \middle| #2 \right)}}

\newcommand*{\veck}{{\mathbf{k}}}
\newcommand*{\vecq}{{\mathbf{q}}}
\newcommand*{\vecG}{{\mathbf{G}}}

\newcommand*{\vecSig}{{\mathbf{\Sigma}}}

\newcommand*{\vecJ}{{\mathbf{J}}}
\newcommand*{\vecr}{{\mathbf{r}}}

\newcommand*{\vecT}{{\mathbf{T}}}

\newcommand*{\vecp}{{\mathbf{p}}}
\newcommand*{\vecI}{{\mathbf{I}}}
\newcommand*{\veczero}{{\mathbf{0}}}
\newcommand*{\vecPi}{{\mathbf{\Pi}}}

\newcommand*{\occ}{{\mathrm{occ}}}

\newcommand*{\vir}{{\mathrm{vir}}}

\definecolor{myblue}{rgb}{0,0,1}
\usepackage[breaklinks=true,colorlinks=true,linkcolor=myblue,urlcolor=myblue,citecolor=myblue]{hyperref}

\author{Tianyu Zhu}
\email{tyzhu@caltech.edu}
\author{Garnet Kin-Lic Chan}
\email{gkc1000@gmail.com}
\affiliation{Division of Chemistry and Chemical Engineering, California Institute of Technology, Pasadena CA 91125}

\title{All-electron Gaussian-based \gw~for Valence and Core Excitation Energies of Periodic Systems}

\begin{tocentry}
\includegraphics[width=0.7\textwidth]{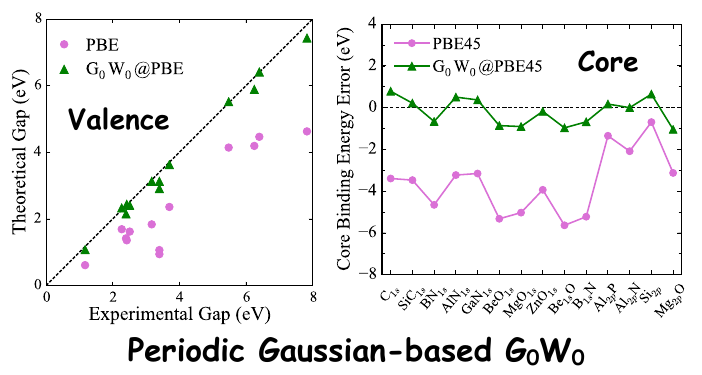}
\end{tocentry}

\begin{document}

\begin{abstract}
  We describe an all-electron \gw~implementation for periodic systems with $\veck$-point sampling implemented
  in a crystalline Gaussian basis. Our full-frequency \gw~method relies on efficient Gaussian density fitting integrals and
  includes both analytic continuation and contour deformation schemes. Due to the compactness of Gaussian bases, no virtual state truncation
  is required as
  is seen in many plane-wave formulations. Finite size corrections are included by taking the $\vecq \to \veczero$ limit of the Coulomb divergence.
  Using our implementation, we study quasiparticle energies and band structures across a range
  of systems including molecules, semiconductors, rare gas solids, and metals.
  We find that the \gw~band gaps of traditional semiconductors converge rapidly with respect to the basis size, even for the conventionally challenging case of ZnO.
  Using 
  correlation-consistent bases of polarized triple-zeta quality, we find the mean absolute relative error of the extrapolated \gw@PBE band gaps
  to be only 5.2\% when compared to experimental values. For core excitation binding energies (CEBEs),
  we find that \gw~predictions improve significantly over those from DFT if the \gw~calculations are started from 
  hybrid functionals with a high percentage of exact exchange.
\end{abstract}

\maketitle

\section{Introduction}
The accurate simulation of materials spectra is an important target of computational solid-state physics and chemistry.
While density functional theory (DFT)~\cite{Kohn1965} has been widely used due to its low computational cost,
the derivative discontinuity means that Kohn-Sham orbital energies do not formally describe the quasiparticle
energies~\cite{Perdew1982,Perdew1983,Sham1985}. In practice, band structures obtained from
the local density approximation (LDA) and generalized gradient approximations (GGA) substantially underestimate band gaps in most
solids~\cite{Perdew2017}. A many-body computation of the quasiparticle energies is thus desirable, and
the \GW~approximation~\cite{Hedin1965,Hanke1979,Strinati1982,Hybertsen1986,Aryasetiawan1998,Golze2019a} is often used for this task. The \GW~approximation
starts from a screened Coulomb interaction ($W$)
treated at the level of the random phase approximation (RPA)~\cite{Ren2012a}, and the method derives
its name from defining the self-energy as $\Sigma = G \ast W$, where $G$ is the one-electron Green's function. 
Because of the simple structure of the self-energy and the associated low computational cost compared to other many-body treatments, \GW~has
become a method of choice for quantitative simulations of quasiparticle spectra in weakly-correlated materials.


Within the \GW~approximation, one can define several levels of self-consistency. Often, \GW~is employed 
in its cheapest form, as a  
non-self-consistent, one-shot, method referred to as \gw. In this case, the accuracy of the method
has a dependence on the starting choice of orbitals and orbital energies which enter into $G_0$~\cite{Bruneval2013,Korzdorfer2012,Chen2014a}. Like other
electronic structure methods, how best to implement \gw~depends on the computational basis.
The most common basis  with which to implement \gw~in periodic systems is the plane-wave (PW) basis~\cite{Gonze2009,Deslippe2012,Marini2009,Govoni2015,Martin-Samos2009,Schlipf2020}. However, this basis has some well-known drawbacks. One is that 
representing the sharp core electron density well requires a large number of plane waves. Thus
either pseudopotentials must be used, or the basis should be augmented by other functions, as is done in muffin tin representations, such as the full-potential linear augmented plane waves (FLAPW)~\cite{Friedrich2010,Jiang2013,gulans2014}, linear muffin-tin orbitals (LMTO)~\cite{Pashov2020}, and projector augmented-wave (PAW)~\cite{Shishkin2006,Liu2016,Gonze2009,Huser2013} techniques.  Pseudopotentials prevent
access to the core excitation binding energies (CEBEs), and while augmented plane wave (APW) schemes in principle allow
core states to be accessed, the implementations are more complicated~\cite{aoki2018,ishii2010}.
A second drawback of plane-wave based \gw~implementations is the need for a large number of virtual states to converge
the screened interaction (even when
computing valence quasiparticle energies~\cite{Golze2019a}), especially for materials with $d$ or $f$ electrons~\cite{Friedrich2011,Stankovski2011,Jiang2018}. Some techniques to alleviate this problem by avoiding explicit sums over virtual states~\cite{Giustino2010,Schlipf2020,Govoni2015,LaflammeJanssen2015} have 
recently appeared.

Local atomic orbital (AO) representations, such as the Gaussian basis sets that are widely used in quantum chemistry methods,
are another potential choice of computational basis. Local AOs are well suited to describe the rapid oscillations of the electron density near nuclei, and treat the core and valence states on an equal footing.
Furthermore, due to the much smaller basis size of typical AO basis sets compared to plane wave bases~\cite{Booth2016},
summations over virtual states can be directly performed without truncation. Several \gw~implementations based on local AO bases, including Gaussians~\cite{Blase2011,VanSetten2013,Bruneval2016,Wilhelm2016,Wilhelm2018,Tirimbo2020} and numeric atom-centered orbitals (NAOs)~\cite{Ren2012,Golze2018a,Koval2019}, have been reported for molecules. However, there are few such implementations for periodic systems. Amongst these, Rohlfing \textit{et al.} carried
out some early explorations using Gaussian orbitals for \gw~within the plasmon-pole model~\cite{Rohlfing1993,Rohlfing1995}, and
recently a $\Gamma$-point only periodic \gw~implementation~using Gaussian basis sets in conjunction with pseudopotentials has been reported~\cite{Wilhelm2017a}. In addition to formulations using plane waves and local AOs, there are also \gw~codes implemented with real space grid bases, such as stochastic \GW~\cite{Vlcek2017} and NanoGW~\cite{Gao2020a}.

In this work, we describe a periodic all-electron full-frequency \gw~implementation using (crystalline) Gaussian basis sets within the PySCF quantum chemistry
platform~\cite{Sun2018,Sun2020}. We note that this implementation was previously used, without a description of the implementation or its performance,
in the context of the full cell \GW+DMFT approach that we have recently developed for strongly correlated solids~\cite{Zhu2020a,Zhu2019,Cui2020,Zhu2020}.
The purpose of this work is to carefully present the algorithm and benchmark its performance.
To this end, we carry out detailed benchmarks for quasiparticle energies, paying attention to basis set convergence and finite size effects,
across a wide range of problems including the excitation energies of molecules, semiconductors, rare gas solids, and metals,
and core excitation binding energies in semiconductors. 
Our \gw~implementation explicitly samples the Brillouin zone (i.e., uses $\veck$-points), and utilizes the efficient periodic Gaussian density fitting integral infrastructure in PySCF~\cite{Sun2017b}.
Different numerical treatments are appropriate for the core and valence excitations, and we
describe and contrast implementations based on the analytic continuation (AC)~\cite{Rojas1995,Rieger1999,Friedrich2010,Liu2016} and contour deformation (CD)~\cite{Godby1988,Gonze2009,Govoni2015} techniques.


\section{Theory and Implementation}
\subsection{\gw~Approximation}
Our periodic \gw~implementation closely follows the molecular \gw~implementations described in Refs.~\cite{Ren2012,Wilhelm2016,Golze2018a}. Assuming a DFT calculation has already been performed on a given periodic system, the key quantity to compute in \gw~is the self-energy
\begin{equation}\label{eq:sigma}
\Sigma(\vecr, \vecr', \omega) = \frac{i}{2\pi} \int_{-\infty}^{\infty} d\omega' e^{i\omega'\eta} G_0(\vecr, \vecr', \omega+\omega') W_0(\vecr, \vecr', \omega') .
\end{equation}
Here, $G_0$ is the non-interacting Green's function, $W_0$ is the screened Coulomb interaction, $\omega$ is the frequency and $\eta$ is a positive infinitesimal. $G_0(\vecr, \vecr', \omega)$ is defined from the DFT (crystalline) molecular orbital (MO) energies $\{\epsilon_{m\veck_m}\}$ and orbitals $\{\psi_{m\veck_m}\}$:
\begin{equation}\label{eq:G0}
G_0(\vecr, \vecr', \omega) = \sum_{m\veck_m} \frac{\psi_{m\veck_m}(\vecr) \psi_{m\veck_m}^*(\vecr')}{\omega-\epsilon_{m\veck_m}-i\eta \mathrm{sgn}(\epsilon_F - \epsilon_{m\veck_m})},
\end{equation}
with $\epsilon_F$ as the Fermi energy and $\veck_m$ as a crystal momentum vector in the first Brillouin zone. The screened Coulomb interaction $W_0(\vecr, \vecr', \omega)$ is defined as
\begin{equation}\label{eq:W0}
W_0(\vecr, \vecr', \omega) = \int d\vecr'' \varepsilon^{-1}(\vecr, \vecr'', \omega) v(\vecr'', \vecr'),
\end{equation}
where $v(\vecr, \vecr') = |\vecr - \vecr'|^{-1}$ is the bare Coulomb operator, and $\varepsilon(\vecr, \vecr', \omega)$ is the dielectric function:
\begin{equation}\label{eq:epsr}
\varepsilon(\vecr, \vecr', \omega) = \delta(\vecr, \vecr') - \int d\vecr'' v(\vecr, \vecr'') \chi_0(\vecr'',\vecr',\omega).
\end{equation}
The polarizability $\chi_0(\vecr,\vecr',\omega)$ is calculated in the random phase approximation (RPA):
\begin{equation}\label{eq:chi}
\begin{split}
\chi_0(\vecr,\vecr',\omega) = \frac{1}{N_\veck} \sum_{i\veck_i}^{\occ} \sum_{a\veck_a}^\vir  \bigg( \frac{\psi_{i\veck_i}^*(\vecr) \psi_{a\veck_a}(\vecr) \psi_{a\veck_a}^*(\vecr') \psi_{i\veck_i}(\vecr')}{\omega - (\epsilon_{a\veck_a} - \epsilon_{i\veck_i}) + i\eta}  - \frac{\psi_{i\veck_i}(\vecr) \psi_{a\veck_a}^*(\vecr) \psi_{a\veck_a}(\vecr') \psi_{i\veck_i}^*(\vecr')}{\omega + (\epsilon_{a\veck_a} - \epsilon_{i\veck_i}) - i\eta} \bigg),
\end{split}
\end{equation}
where $i$ and $a$ label occupied and virtual molecular orbitals respectively.

Once the self-energy is computed, one solves the \gw~quasiparticle (QP) equation to obtain \gw~QP energies: 
\begin{equation}\label{eq:qp}
\epsilon_{n\veck}^{\GW} = \epsilon_{n\veck}^{\mathrm{DFT}} + \left( \psi_{n\veck} \middle| \mathrm{Re} \ \Sigma(\epsilon_{n\veck}^{\GW}) - v^{xc} \middle| \psi_{n\veck} \right) .
\end{equation}
Eq.~\ref{eq:qp} needs to be solved self-consistently. Sometimes an approximate linearization is used:
\begin{equation}\label{eq:linear}
\epsilon_{n\veck}^{\GW} = \epsilon_{n\veck}^{\mathrm{DFT}} + Z_{n\veck} \left( \psi_{n\veck} \middle| \mathrm{Re}\ \Sigma(\epsilon_{n\veck}^{\mathrm{DFT}}) - v^{xc} \middle| \psi_{n\veck} \right),
\end{equation}
where $Z_{n\veck}$ is the renormalization factor
\begin{equation}\label{eq:qpweight}
Z_{n\veck} = \left( 1-\frac{\partial \mathrm{Re} \Sigma_{n\veck}(\omega)}{\partial \omega} \bigg|_{\omega=\epsilon_{n\veck}} \right) ^{-1} .
\end{equation}
In this work, all reported \gw~QP energies are obtained by solving Eq.~\ref{eq:qp} self-consistently using a Newton solver.
Note that in addition to the quasiparticle energies, the full \gw~interacting Green's function can be obtained through Dyson's equation:
\begin{equation}
G^{-1}(\vecr, \vecr', \omega) = G_0^{-1}(\vecr, \vecr', \omega) - \Sigma(\vecr, \vecr', \omega),
\end{equation}
although we do not present results for the full Green's function in this work.

\subsection{Gaussian Density Fitting}
To represent the crystalline molecular orbitals appearing in the formulae above, we employ a single-particle basis of
crystalline Gaussian atomic orbitals. These are translational-symmetry-adapted linear combinations of Gaussian AOs $\{ \tilde{\phi}_\mu \}$:
\begin{equation}
\phi_{\mu \veck} (\vecr) = \sum_{\vecT} e^{i\veck \cdot \vecT} \tilde{\phi}_\mu(\vecr - \vecT),
\end{equation}
where $\vecT$ is a lattice translation vector. 

The most computationally expensive set of matrix elements in this basis to compute and store
are those associated with the electron repulsion integrals (ERIs). Directly computing 4-center ERIs
and using them to construct the \gw~self-energy is particularly expensive in periodic systems.
Therefore, we use the density fitting (resolution of identity) technique~\cite{Whitten1973} to reduce the computational cost of Eqs.~\ref{eq:W0}-\ref{eq:chi} and to avoid storing the full set of 4-center ERIs. We use periodic Gaussian density fitting (GDF) in the Coulomb metric as described in  Ref.~\cite{Sun2017b}, where ERIs are decomposed into 3-center 2-electron integrals: 
\begin{equation}\label{eq:gdf}
\eri{p\veck_p q\veck_q}{r\veck_r s\veck_s} = \sum_{PQ} \eri{p\veck_p q\veck_q}{P \veck_{pq}} \vecJ^{-1}_{PQ} \eri{Q\veck_{rs}}{r\veck_r s\vecr_s},
\end{equation}
where 
\begin{equation}\label{eq:J}
\vecJ_{PQ}(\veck) = \iint d\vecr d\vecr' \phi_{P(-\veck)} (\vecr) \frac{1}{|\vecr - \vecr'|} \phi_{Q\veck} (\vecr'),
\end{equation}
\begin{equation}
\eri{Q\veck_{rs}}{r\veck_r s\veck_s} = \iint d\vecr d\vecr' \phi_{Q\veck_{rs}}(\vecr) \frac{1}{|\vecr - \vecr'|} \phi_{r\veck_r}^* (\vecr') \phi_{s\veck_s} (\vecr') .
\end{equation}
Here, $p, q, r, s$ are crystalline Gaussian AOs and $P, Q$ denote auxiliary periodic Gaussian functions. $\veck_{pq}$, $\veck_p$ and $\veck_q$ satisfy momentum conservation: $\veck_{pq} = \veck_p - \veck_q + N\mathbf{b}$, where $\mathbf{b}$ is a reciprocal lattice vector and $N$ is an integer. $\veck_{pq}$ is restricted to the first Brillouin zone. Because of the momentum conservation relation $\veck_{p} - \veck_q + \veck_r -\veck_s = N\mathbf{b}$, it is also clear that $\veck_{pq} = -\veck_{rs}$. We note that no complex conjugate needs to be used for the auxiliary orbitals in the integrals, since the phases in the crystalline AOs are completely
determined by the crystal momentum. The three-center integral $\eri{Q\veck_{rs}}{r\veck_r s\veck_s}$ is evaluated directly in  $\veck$-space using lattice summations, with a cost of $\mathcal{O}(N_{AO}^2 N_{aux} N_c^2)$, where $N_{AO}$ is the number of AOs, $N_{aux}$ is the number of auxiliary Gaussian functions, and $N_c$ is the number of images in the lattice summation.

By decomposing the inverse Coulomb matrix $\vecJ^{-1}_{PQ}=\sum_{R} \vecJ^{-1/2}_{PR} \vecJ^{-1/2}_{RQ}$, Eq.~\ref{eq:gdf} is further simplified to
\begin{equation}
\eri{p\veck_p q\veck_q}{r\veck_r s\veck_s} = \sum_R v_{R \veck_{pq}}^{p\veck_p, q\veck_q} \cdot v_{R\veck_{rs}}^{r\veck_r, s\veck_s},
\end{equation}
where
\begin{equation}\label{eq:vRrs}
v_{R\veck_{rs}}^{r\veck_r, s\veck_s} = \sum_Q \vecJ^{-\frac{1}{2}}_{RQ}(\veck_{rs}) \cdot \eri{Q\veck_{rs}}{r\veck_r s\veck_s} .
\end{equation}
In practice, $v_{R\veck_{rs}}^{r\veck_r, s\veck_s}$ is computed and stored as $v_{R}^{r\veck_r, s\veck_s}$, because $\veck_{rs}$ is always determined by $\veck_r$ and $\veck_s$.
To maintain consistency with plane-wave expressions in the \gw~literature, we can relabel the $v_{R}^{r\veck_r, s\veck_s}$ integral as $v_{R\vecq}^{r\veck, s\veck-\vecq}$, where $\veck$ and $\vecq$ are crystal momentum vectors. The size of the auxiliary Gaussian basis is normally 3-10 times the size of
the Gaussian AO basis, depending on whether we use an optimized auxiliary basis, or a brute-force even-tempered fitting set. We will describe how to utilize the GDF integrals in the \gw~expressions in the next two sections.

\subsection{Analytic Continuation}\label{sec:AC}
Numerical integration along the real frequency axis, as expressed in Eq.~\ref{eq:sigma}, is challenging because $G_0$ and $W_0$ both
have many poles along the real axis. One way to avoid this problem is to perform the integration along the
imaginary frequency axis~\cite{Ren2012,Wilhelm2016}: 
\begin{equation}
\Sigma(\vecr, \vecr', i\omega) = -\frac{1}{2\pi} \int_{-\infty}^{\infty} d\omega' G_0(\vecr, \vecr', i\omega+i\omega') W_0(\vecr, \vecr', i\omega'),
\end{equation}
and then to analytically continue the self-energy to the real axis to computing QP energies and other spectral quantities.

The non-interacting Green's function $G_0$ on the imaginary axis becomes
\begin{equation}\label{eq:G0AC}
G_0(\vecr, \vecr', i\omega) = \sum_{m\veck_m} \frac{\psi_{m\veck_m}(\vecr) \psi_{m\veck_m}^*(\vecr')}{i\omega + \epsilon_F - \epsilon_{m\veck_m}}.
\end{equation}
For gapped systems, we take the Fermi energy $\epsilon_F$ to be the midpoint between the DFT valence band maximum and conduction band minimum energies.

The self-energy matrix elements in the MO basis are then computed as
\begin{equation}\label{eq:sigmaAC}
\begin{split}
\vecSig_{nn'}(\veck, i\omega) =& -\frac{1}{2\pi N_\veck} \sum_{m\vecq} \int_{-\infty}^{\infty} d\omega' [\vecG_0(\veck-\vecq, i\omega+i\omega')]_{mm} \\
&  \times \left( n\veck, m\veck-\vecq \middle| W_0 \middle| m\veck-\vecq, n'\veck \right) .
\end{split}
\end{equation}
Here, $N_\veck$ is the number of sampled $\veck$-points, indices $n$, $n'$ and $m$ refer to molecular orbitals. 
Molecular orbitals $n$ and $n'$ share the same $\veck$ and $m$ appears with $\veck-\vecq$.
To compute the matrix elements of $W_0$ in Eq.~\ref{eq:sigmaAC}, we rewrite Eqs.~\ref{eq:W0} and \ref{eq:epsr} as an infinite summation by a Taylor expansion:
\begin{equation}\label{eq:W0inf}
W_0 = v + v\chi_0 v + v\chi_0 v \chi_0 v + \ldots~,
\end{equation}
where $v\chi_0 v$ and $v\chi_0 v \chi_0 v$ involve integrations over real-space coordinates. Meanwhile, we expand the orbital pair product $\psi_{n\veck}^*(\vecr) \psi_{m\veck-\vecq}(\vecr)$ in the auxiliary basis
\begin{equation}\label{eq:orbpair}
\psi_{n\veck}^*(\vecr) \psi_{m\veck-\vecq}(\vecr) = \sum_P b_{P\vecq}^{n\veck, m\veck-\vecq} \phi_{P\vecq} (\vecr),
\end{equation}
with
\begin{equation}\label{eq:bPnm}
b_{P\vecq}^{n\veck, m\veck-\vecq} = \sum_R (n\veck, m\veck-\vecq | R\vecq) \cdot \vecJ^{-1}_{RP}(\vecq).
\end{equation}
To ease notation, some momentum labels are suppressed in the above and following equations (e.g., we will use $b_P^{nm}$ to denote $b_{P\vecq}^{n\veck, m\veck-\vecq}$). Using Eqs.~\ref{eq:W0inf}-\ref{eq:bPnm}, the matrix elements of $W_0$ are computed as
\begin{equation}\label{eq:Wmat}
\begin{split}
& \left( n\veck, m\veck-\vecq \middle| W_0 \middle| m\veck-\vecq, n'\veck \right) \\
&= \sum_{PQ} b_P^{nm} \big[ \iint d\vecr d\vecr' \phi_{P\vecq}(\vecr) W_0(\vecr,\vecr',i\omega') \phi_{Q(-\vecq)}(\vecr') \big] b_Q^{mn'} \\
&= \sum_{PQ} b_P^{nm} \big[ \vecJ_{PQ}(\vecq) + (\vecJ^{1/2} \vecPi \vecJ^{1/2})_{PQ}(\vecq) + ... \big]  b_Q^{mn'} \\
&= \sum_{PQ} v_P^{nm} [ \vecI-\vecPi(\vecq, i\omega') ]_{PQ}^{-1} v_Q^{mn'}.
\end{split}
\end{equation}
The 3-center 2-electron integral $v_P^{nm}$ between auxiliary basis function $P$ and molecular orbital pairs $nm$ is
obtained from an AO to MO transformation of the GDF AO integrals defined in Eq.~\ref{eq:vRrs}:
\begin{equation}\label{eq:gdfmo}
v_P^{nm} = \sum_p\sum_q C_{pn}(\veck) C_{qm}(\veck-\vecq) v_{P\vecq}^{p\veck, q\veck-\vecq},
\end{equation}
where $C(\veck)$ refers to the MO coefficients in the AO basis. $\vecPi(\vecq, i\omega')$ in Eq.~\ref{eq:Wmat} is an auxiliary density response function:
\begin{equation}\label{eq:Pi}
\begin{split}
\vecPi_{PQ}(\vecq, i\omega') = \frac{2}{N_\veck} \sum_{\veck} \sum_{i}^{\occ}\sum_{a}^{\vir} v_{P}^{ia} \frac{\epsilon_{i\veck} - \epsilon_{a\veck-\vecq}}{\omega'^2 + (\epsilon_{i\veck} - \epsilon_{a\veck-\vecq})^2} v_{Q}^{ai} .
\end{split}
\end{equation}

Inserting Eq.~\ref{eq:G0AC} and Eq.~\ref{eq:Wmat}, Eq.~\ref{eq:sigmaAC} becomes 
\begin{equation}\label{eq:sigmaAC2}
\begin{split}
\vecSig_{nn'}(\veck, i\omega) = & -\frac{1}{2\pi N_\veck} \sum_{m\vecq} \int_{-\infty}^{\infty} d\omega' \frac{1}{i(\omega+\omega')+\epsilon_F-\epsilon_{m\veck-\vecq}}  \\
&  \times \sum_{PQ} v_P^{nm} [\vecI-\vecPi(\vecq, i\omega')]_{PQ}^{-1} v_Q^{mn'} .
\end{split}
\end{equation}

The self-energy term in Eq.~\ref{eq:sigmaAC2} is further divided into exchange and correlation components $\vecSig(\veck, i\omega) = \vecSig^x(\veck) + \vecSig^c(\veck, i\omega)$, where the frequency-independent exchange $\vecSig^x(\veck)$ is the Hartree-Fock (HF) exchange matrix evaluated using the DFT orbitals: 
\begin{equation}\label{eq:sigmax}
\vecSig_{nn'}^x(\veck) = -\frac{1}{N_\veck} \sum_{P\vecq} \sum_{i}^{\occ} v_{P\vecq}^{n\veck, i\veck-\vecq} \cdot v_{P(-\vecq)}^{i\veck-\vecq, n'\veck} .
\end{equation}
The advantage of this division is that the HF exchange is free of integration error. Accordingly, the correlation part of self-energy becomes
\begin{equation}\label{eq:sigmaAC3}
\begin{split}
\vecSig^c_{nn'}(\veck, i\omega) = & -\frac{1}{\pi N_\veck} \sum_{m\vecq} \int_{0}^{\infty} d\omega' \frac{i\omega+\epsilon_F-\epsilon_{m\veck-\vecq}}{(i\omega+\epsilon_F-\epsilon_{m\veck-\vecq})^2 + \omega'^2}  \\
&  \times \sum_{PQ} v_P^{nm} \big[ [\vecI-\vecPi(\vecq, i\omega')]^{-1}_{PQ}-\delta_{PQ} \big] v_Q^{mn'} .
\end{split}
\end{equation}

The integration in Eq.~\ref{eq:sigmaAC3} must be performed on a numerical grid. Following Ref.~\cite{Ren2012}, we employ a modified Gauss-Legendre grid that transforms a standard Gauss-Legendre grid in the range $[-1, 1]$ to $[0, \infty]$:
\begin{equation}
\tilde{x_i} = \frac{1+x_i}{2(1-x_i)}, ~~\tilde{w_i} = \frac{w_i}{(1-x_i)^2}.
\end{equation}
$\{x_i, w_i\}$  are the original Gauss-Legendre abscissas and weights of the grid points in the range $[-1, 1]$, and $\{\tilde{x}_i, \tilde{w}_i\}$ are the transformed abscissas and weights of the grid points used for integrating Eq.~\ref{eq:sigmaAC3}. In this work, $N_G =100$ grid points were used for all reported results unless specified.

The real-frequency self-energy is then obtained through an analytic continuation scheme by fitting to an analytic function defined
on the full complex plane. We fit the self-energy matrix elements to $N$-point Pad\'e approximants using Thiele's reciprocal difference method (see Appendix of Ref.~\cite{Vidberg1977} for the detailed algorithm):
\begin{equation}\label{eq:pade}
\vecSig_{nn'}(\veck,z) = \frac{a_0+a_1 \cdot z+\ldots+a_{(N-1)/2} \cdot z^{(N-1)/2}}{1+b_1 \cdot z+\ldots+b_{N/2} \cdot z^{N/2}} ,
 \end{equation}
where $z$ is any complex frequency, $\{a_i, b_i\}$ are Pad\'e coefficients that need to be fitted. In this work, we use $N=18$ Pad\'e approximants, which requires 18 $\vecSig_{nn'}(\veck,z=i\omega)$ data points as input to the fit. We choose these 18 imaginary frequency points to be the positions
of the modified Gauss-Legendre grid points. Once the Pad\'e polynomials are fitted, the real-frequency self-energy is easily calculated:
\begin{equation}\label{eq:pade2}
\vecSig_{nn'}(\veck,\omega) =  \frac{a_0+a_1 \cdot \omega+\ldots+a_{(N-1)/2} \cdot \omega^{(N-1)/2}}{1+b_1 \cdot \omega+\ldots+b_{N/2} \cdot \omega^{N/2}}.
 \end{equation}

We further comment on the computational scaling of this approach. If only \gw~QP energies are required, one only needs to compute the diagonal self-energy matrix elements. The most expensive step is then computing the auxiliary density response function in Eq.~\ref{eq:Pi}, whose cost scales as $\mathcal{O}(N_\veck^2 N_{o} N_{v} N_{aux}^2)$, where $N_o, N_v, N_{aux}$ are the number of occupied, virtual and auxiliary orbitals per unit cell. There is also a prefactor $N_G$ (number of grid points) in Eq.~\ref{eq:Pi}, which is fixed regardless of the system size. The second expensive step is integrating Eq.~\ref{eq:sigmaAC3}, which scales as $\mathcal{O}(N_\veck^2 N_{AO} N_{aux}^2)$ and has a prefactor $N_G N_{Pade}$, with $N_{AO}$ as the number of atomic orbitals per unit cell and $N_{Pade}$ as the number of Pad\'e approximants. On the other hand, if the full \gw~Green's function and off-diagonal self-energy matrix at all $\veck$-points are also required, Eq.~\ref{eq:sigmaAC3} becomes the most time-consuming step, with a cost scaling of $\mathcal{O}(N_\veck^2 N_{AO}^2 N_{aux}^2)$.

\subsection{Contour Deformation}
\begin{figure}[hbt]
\centering
\includegraphics[width=2.5in]{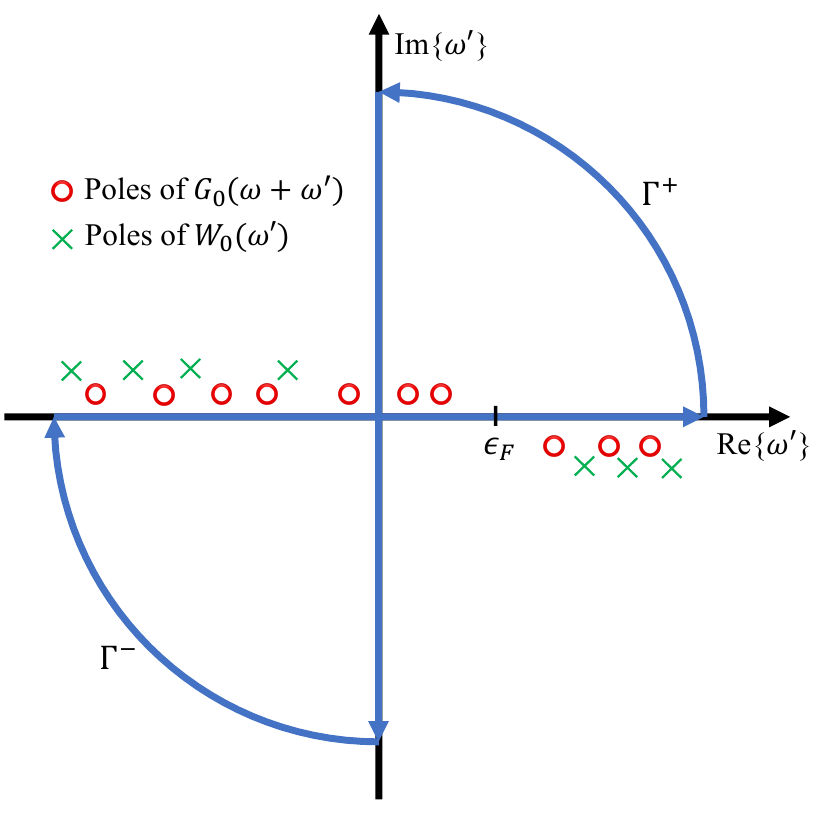}
\caption{Contours used for integration in the \gw-CD approach. The integration contours $\Gamma^+$ and $\Gamma^-$ enclose some poles of the Green's function $G_0$, but exclude poles of the screened Coulomb interaction $W_0$.}
\label{fig:cd}
\end{figure}

The AC scheme has been shown to give accurate \gw~valence state energies in molecules and solids in general, although in some systems a higher Pad\'e expansion (e.g., greater than 100) is needed to properly converge the quasiparticle energies~\cite{VanSetten2015,Wilhelm2016,Liu2016}. However, analytic continuation is known to be very unstable for states far away from the Fermi level (e.g., core excitation energies) and thus gives inaccurate \gw~QP energies for those states~\cite{Duchemin2020}. A more robust scheme is the contour deformation (CD) approach~\cite{Golze2018a,Govoni2015}, where the integral over the real frequency axis in Eq.~\ref{eq:sigma} is transformed into an integral over the contours shown in Fig.~\ref{fig:cd}.

The integration in Eq.~\ref{eq:sigma} can then be broken down into the sum of an integration along the imaginary frequency axis
plus all residues arising from poles enclosed within the contours:
\begin{equation}\label{eq:sigmaCD}
\begin{split}
\Sigma(\vecr, \vecr', \omega) &= \frac{i}{2\pi} \oint d\omega' G_0(\vecr, \vecr', \omega+\omega') W_0(\vecr, \vecr', \omega') \\
& ~~~~ -\frac{1}{2\pi} \int_{-\infty}^{\infty} d\omega' G_0(\vecr, \vecr', \omega+i\omega') W_0(\vecr, \vecr', i\omega') \\
& = \Sigma^{C}(\vecr, \vecr', \omega) + \Sigma^I(\vecr, \vecr', \omega).
\end{split}
\end{equation}
From Eq.~\ref{eq:G0}, the poles of $G_0(\vecr, \vecr', \omega+\omega')$ are located at frequencies
\begin{equation}\label{eq:G0pole}
\omega_{m\veck_m}' = \epsilon_{m\veck_m} - \omega + i\eta \mathrm{sgn} (\epsilon_F - \epsilon_{m\veck_m}),
\end{equation}
with residues
\begin{equation}\label{eq:res}
\mathrm{Res} \{ G_0(\vecr, \vecr', \omega+\omega'), \omega_{m\veck_m}' \} = \psi_{m\veck_m}(\vecr) \psi_{m\veck_m}^*(\vecr') .
\end{equation}
The poles of $G_0(\vecr, \vecr', \omega+\omega')$ thus enter the contour $\Gamma^+$ when $\epsilon_{m\veck_m}<\epsilon_F$ and $\omega_{m\veck_m}'>0$. Similarly, when $\epsilon_{m\veck_m}>\epsilon_F$ and $\omega_{m\veck_m}'<0$, the poles of $G_0(\vecr, \vecr', \omega+\omega')$ enter the contour $\Gamma^-$. On the other hand, the poles of $W_0(\vecr, \vecr', \omega')$ are never enclosed in the contours in Fig.~\ref{fig:cd}. Therefore, the integral of the contour in Eq.~\ref{eq:sigmaCD} is computed as
\begin{equation}\label{eq:sigmaCD2}
\begin{split}
\Sigma^{C}_{nn'}(\veck, \omega) &= \frac{1}{N_\veck} \sum_{m\vecq} f_{m\veck-\vecq} 
 \left( n\veck, m\veck-\vecq \middle| W(\omega_{m\veck-\vecq}') \middle| m\veck-\vecq, n'\veck \right) \\
&= \frac{1}{N_\veck} \sum_{m\vecq} f_{m\veck-\vecq}
\sum_{PQ} v_P^{nm} [\vecI-\vecPi(\vecq, \omega_{m\veck-\vecq}')]_{PQ}^{-1} v_Q^{mn'} ,
\end{split}
\end{equation}
 and the contribution of residues $f_{m\veck-\vecq}$ is given by
\begin{equation}
f_{m\veck-\vecq} = 
    \begin{cases}
      ~1 &  \text{if} ~\epsilon_F < \epsilon_{m\veck-\vecq} < \omega ;  \\ 
      -1 & \text{if} ~\epsilon_F > \epsilon_{m\veck-\vecq} > \omega ; \\
     ~ 0 & \text{else}.
    \end{cases}
\end{equation}
The auxiliary density response function is computed as (we use $\eta=0.001$ a.u.)
\begin{equation}
\begin{split}
\vecPi_{PQ}(\vecq,\omega) = \frac{1}{N_\veck} \sum_{\veck} \sum_{i}^{\occ}  \sum_{a}^\vir  v_{P}^{ia} \bigg( \frac{1}{\omega - (\epsilon_{a\veck-\vecq} - \epsilon_{i\veck}) + i\eta}
 - \frac{1}{\omega + (\epsilon_{a\veck-\vecq} - \epsilon_{i\veck}) - i\eta} \bigg) v_Q^{ai} .
\end{split}
\end{equation}

The other integration over the imaginary frequency axis in Eq.~\ref{eq:sigmaCD} is calculated on the modified Gauss-Legendre grid, as described in Section~\ref{sec:AC}:
\begin{equation}
\begin{split}\label{eq:sigmaCD3}
\vecSig_{nn'}^I(\veck, \omega) = & -\frac{1}{2\pi N_\veck} \sum_{m\veck-\vecq} \int_{-\infty}^{\infty} d\omega' \frac{1}{\omega+i\omega'+\epsilon_F-\epsilon_{m\veck-\vecq}}  \\
&  \times \sum_{PQ} v_P^{nm} [\vecI-\vecPi(\vecq, i\omega')]_{PQ}^{-1} v_Q^{mn'} ,
\end{split}
\end{equation}
where the expression of $\vecPi(\vecq, i\omega')$ follows Eq.~\ref{eq:Pi}. We note that similar to Eqs.~\ref{eq:sigmax} and \ref{eq:sigmaAC3}, we also separately compute the exchange and correlation parts of self-energy in Eqs.~\ref{eq:sigmaCD2} and \ref{eq:sigmaCD3}.

The \gw-CD approach has the advantage that the real-axis self-energy is directly computed without the need for analytic continuation. However, the computational expense is higher than in the \gw-AC approach. It is clear that computing the imaginary integration alone in Eq.~\ref{eq:sigmaCD3}  has a similar computational cost to the \gw-AC scheme. The extra cost comes from computing the pole residues in Eq.~\ref{eq:sigmaCD2}, which become more expensive if more poles need to be calculated (i.e., if the targeted states are further away from the Fermi level). An extreme case is the core excitations, where the computational scaling of Eq.~\ref{eq:sigmaCD2} is approximately $\mathcal{O}(N_\veck^3 N_{o}^2 N_{v} N_{aux}^2)$. Nevertheless, as shown in previous literature, the \gw-CD scheme needs to be chosen over the \gw-AC scheme in order to obtain numerically stable core excitation energies in molecules~\cite{Golze2018a,Golze2020}. 

\subsection{Coulomb Divergence Correction}\label{sec:head}
The periodic GDF Coulomb integral $v_{R\vecq}^{r\veck,s\veck-\vecq}$ removes the $\vecG=\veczero$ contribution (here $\vecG$ is a reciprocal lattice vector) when $\vecq=\veczero$, because the Coulomb interaction $4\pi/\Omega \vecG^2$ diverges at $\vecG=\veczero$. This leads to an $\mathcal{O}(N_\veck^{-1/3})$ finite size error in our \gw~calculations. To correct the leading order finite size error
in the correlation and exchange self-energies, we rewrite the \gw-AC self-energy expression in Eq.~\ref{eq:sigmaAC3} in a plane-wave expansion~\cite{Huser2013}:
\begin{equation}\label{eq:sigmaPW}
\begin{split}
\vecSig^c_{nn'}(\veck, i\omega) = -\frac{1}{2\pi N_\veck \Omega_\mathrm{cell}} & \sum_{m\vecq} \sum_{\vecG \vecG'} \int_{-\infty}^{\infty} d\omega' W_{\vecG \vecG'}(\vecq, i\omega') \\
& \times \frac{\rho_{m\veck-\vecq}^{n\veck}(\vecG) \rho_{m\veck-\vecq}^{n'\veck*}(\vecG')}{i(\omega+\omega')+\epsilon_F-\epsilon_{m\veck-\vecq}} ,
\end{split}
\end{equation}
where $\Omega_\mathrm{cell}$ is the unit cell volume. The pair density $nm$ is now expanded in plane waves ($\vecG$ and $\vecG'$), with the matrix elements
\begin{equation}\label{eq:rhoG}
\rho_{m\veck-\vecq}^{n\veck}(\vecG) = (n\veck | e^{i(\vecq+\vecG)\vecr} | m\veck-\vecq ).
 \end{equation}
The screened Coulomb potential is computed as:
\begin{equation}\label{eq:WG}
W_{\vecG \vecG'}(\vecq,i\omega) = \frac{\sqrt{4\pi}}{|\vecq+\vecG|} \big(\epsilon_{\vecG \vecG'}^{-1}(\vecq, i\omega) - \delta_{\vecG \vecG'} \big) \frac{\sqrt{4\pi}}{|\vecq+\vecG'|}, 
 \end{equation}
with the dielectric function defined as:
\begin{equation}\label{eq:epsG}
\epsilon_{\vecG \vecG'}(\vecq, i\omega) = \delta_{\vecG \vecG'} - \frac{\sqrt{4\pi}}{|\vecq+\vecG|} \chi_{\vecG \vecG'}(\vecq, i\omega) \frac{\sqrt{4\pi}}{|\vecq+\vecG'|},
 \end{equation}
and the polarizability kernel is computed as:
\begin{equation}\label{eq:chiG}
\begin{split}
\chi_{\vecG \vecG'}(\vecq, i\omega) = & \frac{2}{N_\veck \Omega_\mathrm{cell}}  \sum_{\veck} \sum_{i}^{\occ}\sum_{a}^{\vir} \rho_{a\veck-\vecq}^{i\veck}(\vecG) \\ 
& \times \frac{\epsilon_{i\veck} - \epsilon_{a\veck-\vecq}}{\omega^2 + (\epsilon_{i\veck} - \epsilon_{a\veck-\vecq})^2} \rho_{a\veck-\vecq}^{i\veck *}(\vecG') .
\end{split}
\end{equation}
In Eq.~\ref{eq:chiG}, when $\vecG=\veczero$, $\rho_{a\veck-\vecq}^{i\veck}(\vecG) \approx i\vecq \cdot (i\veck | \vecr | a\veck-\vecq )$ at $\vecq \rightarrow \veczero$, so $\chi_{\veczero \veczero}(\vecq \rightarrow \veczero, i\omega)=\mathcal{O}(\vecq^2)$. Therefore, in Eq.~\ref{eq:epsG}, $\epsilon_{\veczero \veczero}(\vecq \rightarrow \veczero, i\omega)$ has finite value. This means the head of the screened Coulomb potential $W_{\veczero \veczero}$ diverges as $\mathcal{O}(1/\vecq^2)$ when $\vecq \rightarrow \veczero$. Similarly, the wings of the screened Coulomb potential $W_{\vecG \veczero}$ and $W_{\veczero \vecG'}$ diverge as $\mathcal{O}(1/\vecq)$. However, in the limit of a very fine $\veck$-point sampling, $\sum_\vecq \rightarrow \frac{\Omega}{(2\pi)^3}\int dq 4\pi q^2$ ($\Omega=\Omega_\mathrm{cell} N_\veck$), thus $W_{\veczero \veczero}$ and $W_{\vecG \veczero}$ are integrable. 

Following Refs.~\cite{Huser2013,Jiang2013,Wilhelm2017a}, we add head and wings finite size corrections in our \gw~implementation.
We determine the contributions for $\vecq=\veczero, \vecG=\veczero$ by analytically integrating the $\vecq\to \veczero$ contributions in a sphere of radius $\vecq=q_0$ around $\vecG=\veczero$ ($\frac{\Omega}{(2\pi)^3}\int_0^{q_0} dq 4\pi q^2=1$). 
Doing this for the head of screened Coulomb potential in Eq.~\ref{eq:WG} gives 
\begin{equation}\label{eq:Whead}
W_{\veczero \veczero}(\vecq=\veczero,i\omega) = \frac{2\Omega}{\pi} \left(\frac{6\pi^2}{\Omega} \right)^{1/3} \big[\epsilon^{-1}_{\veczero \veczero}(\vecq=\veczero,i\omega)-1 \big].
 \end{equation}
In the $\vecq \rightarrow \veczero$ limit, $\rho_{m\veck-\vecq}^{n\veck}(\veczero) \approx 1$ only at $n=m$. Therefore, inserting Eq.~\ref{eq:Whead} into Eq.~\ref{eq:sigmaPW}, one arrives at the head correction to the self-energy (which only modifies the diagonal self-energy matrix elements):
\begin{equation}\label{eq:sigmahead}
\begin{split}
\vecSig^{\mathrm{head}}_{nn}(\veck, i\omega) = & -\frac{1}{\pi^2} \left(\frac{6\pi^2}{\Omega_\mathrm{cell} N_\veck} \right)^{1/3} \\
& \times \int_{-\infty}^{\infty} d\omega' \frac{\epsilon^{-1}_{\veczero \veczero}(\vecq = \veczero,i\omega)-1}{i(\omega+\omega')+\epsilon_F-\epsilon_{n\veck}} ,
\end{split}
\end{equation}
where we assume $\epsilon_{n\veck-\vecq}=\epsilon_{n\veck}$. From Eq.~\ref{eq:sigmahead} one can clearly see that ignoring this term leads to an $\mathcal{O}(N_\veck^{-1/3})$ finite size error. Similarly, we can obtain the wings contribution to the self-energy:
\begin{equation}\label{eq:sigmawings}
\begin{split}
\vecSig^{\mathrm{wings}}_{nn'}(\veck, i\omega) = & -\frac{1}{\pi} \sqrt{\frac{\Omega_\mathrm{cell}}{4\pi^3}} \left(\frac{6\pi^2}{\Omega_\mathrm{cell} N_\veck} \right)^{2/3} \\
& \times \sum_P \int_{-\infty}^{\infty} d\omega' \frac{\mathrm{Re} \big[v_P^{nn'} \epsilon^{-1}_{P \veczero}(\vecq= \veczero,i\omega) \big]}{i(\omega+\omega')+\epsilon_F-\epsilon_{n'\veck}} .
\end{split}
\end{equation}
Here, $P$ refers to the Gaussian auxiliary basis and $v_P^{nn'}$ is the GDF integral defined as $v_{P\veczero}^{n\veck,n'\veck}$. The $\vecG=\veczero$ Coulomb divergence correction to the exchange self-energy is derived similarly: 
\begin{equation}\label{eq:excorr}
\vecSig^{x}_{ii}(\veck,i\omega) = -\frac{2}{\pi} \left(\frac{6\pi^2}{\Omega_\mathrm{cell} N_\veck} \right)^{1/3},
 \end{equation}
which is a constant that only applies to the diagonal occupied elements of the exchange self-energy.

The head and wings of the dielectric function in the $\vecq= \veczero$ limit are computed as 
\begin{equation}\label{eq:eps00}
h_{\veczero \veczero}(\vecq= \veczero, i\omega) = 1-\lim_{\vecq\to\veczero} \frac{4\pi}{|\vecq|^2} \chi_{\veczero \veczero}(\vecq \rightarrow \veczero, i\omega) ,
 \end{equation}
\begin{equation}\label{eq:epsP0}
w_{P \veczero}(\vecq= \veczero, i\omega) = \lim_{\vecq\to \veczero} \frac{\sqrt{4\pi}}{|\vecq|} \chi_{P \veczero}(\vecq \rightarrow \veczero, i\omega) ,
\end{equation}
where the polarizability is
\begin{equation}\label{eq:chi00}
\begin{split}
\chi_{\veczero \veczero}(\vecq \rightarrow \veczero, i\omega) = & \frac{2}{N_\veck \Omega_\mathrm{cell}}  \sum_{\veck} \sum_{i}^{\occ}\sum_{a}^{\vir} \rho_{a\veck-\vecq}^{i\veck}(\veczero) \\ 
& \times \frac{\epsilon_{i\veck} - \epsilon_{a\veck}}{\omega^2 + (\epsilon_{i\veck} - \epsilon_{a\veck})^2} \rho_{a\veck-\vecq}^{i\veck *}(\veczero) ,
\end{split}
\end{equation}
\begin{equation}\label{eq:chiP0}
\begin{split}
\chi_{P \veczero}(\vecq \rightarrow \veczero, i\omega) = & \frac{2}{N_\veck \Omega_\mathrm{cell}^{1/2}}  \sum_{\veck} \sum_{i}^{\occ}\sum_{a}^{\vir} v_P^{ia} \\ 
& \times \frac{\epsilon_{i\veck} - \epsilon_{a\veck}}{\omega^2 + (\epsilon_{i\veck} - \epsilon_{a\veck})^2} \rho_{a\veck-\vecq}^{i\veck *}(\veczero) .
\end{split}
\end{equation}
In Eqs.~\ref{eq:eps00} and \ref{eq:epsP0} we evaluate the limit using $\vecq=(0.001,0,0)$; we neglect the anisotropy of this limit
for simplicity in this work, although anisotropic corrections can be obtained as discussed in Ref.~\cite{Freysoldt2007}.
The pair density matrix in the long-wavelength limit is computed using $\veck \cdot \vecp$ perturbation theory as described in Ref.~\cite{Yan2011}:
\begin{equation}\label{eq:kp}
\rho_{a\veck-\vecq}^{i\veck}(\veczero) \big|_{\vecq \rightarrow \veczero} = \frac{-i\vecq \cdot ( \psi_{i\veck} | \nabla | \psi_{a\veck})}{\epsilon_{a\veck}-\epsilon_{i\veck}}.
\end{equation}

Finally, we note that $\epsilon^{-1}_{\veczero \veczero}(\vecq= \veczero,i\omega)$ and $\epsilon^{-1}_{P \veczero}(\vecq = \veczero,i\omega)$ in Eqs.~\ref{eq:sigmahead} and \ref{eq:sigmawings} are not simple inverses of $h_{\veczero \veczero}$ and $w_{P \veczero}$ in Eqs.~\ref{eq:eps00} and \ref{eq:epsP0}. Instead, they are matrix elements of the inverse of the full dielectric matrix~\cite{Jiang2013,Wilhelm2017a}:
\begin{equation}\label{eq:fulleps}
\epsilon_\mathrm{full}(\vecq = \veczero,i\omega) = 
\begin{bmatrix} h_{\veczero \veczero} & w^{\dagger}_{P \veczero} \\ w_{P \veczero} & B_{PQ} \end{bmatrix},
\end{equation}
where $B_{PQ}=[\vecI-\vecPi]_{PQ}$ is the body of the dielectric function computed in the Gaussian auxiliary basis (see Eq.~\ref{eq:Pi}). By inverting Eq.~\ref{eq:fulleps}, one obtains $\epsilon^{-1}_{\veczero \veczero}(\vecq= \veczero,i\omega)$ and $\epsilon^{-1}_{P \veczero}(\vecq= \veczero,i\omega)$ as
\begin{equation}
\epsilon^{-1}_{\veczero \veczero} = 1/ \big( h_{\veczero \veczero} - \sum_{PQ} w^\dagger_{P\veczero} B_{PQ}^{-1} w_{Q\veczero} \big),
\end{equation}
\begin{equation}
\epsilon^{-1}_{P \veczero} = -\epsilon^{-1}_{\veczero \veczero} \sum_{Q} B_{PQ}^{-1} w_{Q\veczero} .
\end{equation}
The Coulomb divergence corrections in the \gw-CD approach are also implemented in a similar manner.

\section{\gw~Benchmark Results}
We benchmark our Gaussian-based \gw~method, as implemented in PySCF, on both molecules and periodic crystals. We use standard all-electron Gaussian basis sets (all Gaussian
basis sets can be found in the Basis Set Exchange database~\cite{Pritchard2019}), and we include all electrons in the \gw~calculations. We note that our
implementation can also be used in Gaussian bases with pseudopotentials~\cite{Goedecker96} or effective core potentials~\cite{Hay1985}, although we do
not present such calculations in this work.

\subsection{Validation of \gw-AC Code}
We first validate our \gw-AC code for molecules (i.e., without periodic boundary conditions) drawn from the \GW100 test set~\cite{VanSetten2015}. The GW100 results have previously been well reproduced by plane-wave and PAW implementations~\cite{Maggio2017,Govoni2018}. Ionization potentials (IP) and QP lowest molecular orbital energies (LUMO) were calculated for 18 molecules from the test set. Following Ref.~\cite{VanSetten2015}, we used the def2-QZVP basis~\cite{Weigend2005} as the orbital basis and the def2-QZVP-RIFIT basis~\cite{Hattig2005} as the auxiliary basis, and the PBE density functional~\cite{Perdew96PBE} was chosen to provide the starting choice of DFT orbitals and energies. Our \gw@PBE results are shown in Table~\ref{tab:gw100}, and compared to those results listed as AIMS-P16 and TM-RI in Ref.~\cite{VanSetten2015} (note
  the TM-RI implementation also used Gaussian density fitting ERIs). As shown in the table, our \gw@PBE results are in very good agreement with
the reference data, confirming the accuracy of our implementation for molecules.

\begin{table}[hbt]
	\centering
	\caption{\gw@PBE IP and QP-LUMO energies of 18 molecules from the \GW100 set~\cite{VanSetten2015} in the def2-QZVP basis using the \gw-AC approach. All numbers are in eV. MAD stands for mean absolute difference between two methods. The \gw~energies of O\textsubscript{3}, BeO, and MgO are compared against the boldface values in Table 4 of Ref.~\cite{VanSetten2015}.} 
	\label{tab:gw100}
	\begin{tabular}{>{\centering\arraybackslash}p{5.5cm}>{\centering\arraybackslash}p{1.7cm}>{\centering\arraybackslash}p{2cm}}
	\hline\hline
Molecule	 &  IP  &  QP-LUMO   \\
	\hline
CH\textsubscript{4}   &   13.93  &  2.45    \\
H\textsubscript{2}O &  11.97 & 2.37   \\
SiH\textsubscript{4} & 12.31 & 2.51  \\
LiH & 6.52 & -0.07 \\
CO & 13.57 & 0.67 \\
CO\textsubscript{2} & 13.25 & 2.50  \\
SO\textsubscript{2} & 11.82 & -1.00  \\
N\textsubscript{2} & 14.89 & 2.45 \\
P\textsubscript{2} & 10.21 & -0.72 \\
Cu\textsubscript{2} & 7.52 & -0.96 \\
NaCl & 8.10 & -0.39 \\
BrK& 7.33 & -0.31 \\
TiF\textsubscript{4} & 13.90 & -0.60 \\
C\textsubscript{6}H\textsubscript{6} & 8.99 & 1.09 \\
C\textsubscript{5}H\textsubscript{5}N\textsubscript{5}O (guanine)  & 7.69 & 0.75 \\
O\textsubscript{3}& 11.61 & -2.30 \\
BeO& 9.47 & -2.49 \\
MgO& 6.78 & -1.90 \\
\hline
MAD (PySCF $-$ AIMS-P16~\cite{VanSetten2015}) & \textbf{0.03} & \textbf{0.01} \\
MAD (PySCF $-$ TM-RI~\cite{VanSetten2015}) & \textbf{0.07} & \textbf{0.03} \\
    \hline\hline
	\end{tabular}
\end{table}

\subsection{AC vs. CD Schemes}\label{sec:accd}
We next apply our periodic \gw-AC and \gw-CD implementations to 11 prototypical semiconductors (Si and C in the diamond structure; SiC, AlP, BN, BP, GaN, GaP, ZnO and ZnS in the zinc blende structure; MgO in the rock salt strcuture) and 2 rare gas solids (Ne and Ar in the fcc structure).
The aim here is to compare the accuracy of \gw-AC against the more robust \gw-CD method for obtaining valence state energies, rather than obtaining converged band gap results (which will be presented in Section~\ref{sec:bandgap}). Thus, here we used Dunning's correlation consistent cc-pVTZ basis~\cite{Dunning1989,Woon1993,Wilson1999,Prascher2011,Balabanov2005} (cc-pVTZ-RI~\cite{Weigend2002a,Hattig2005,Hill2008} was used as the auxiliary basis) and a moderate $4\times4\times4$ $\veck$-mesh for all the tested systems. For MgO, the cc-pVTZ basis exhibits severe linear dependencies, so cc-pVDZ/cc-pVDZ-RI basis sets were used instead. Another possible way to remove linear dependencies in the Gaussian basis is to diagonalize the overlap matrix and remove the eigenvectors associated with small eigenvalues below a threshold. For the rare gas solids, aug-cc-pVTZ/aug-cc-pVTZ-RI basis sets~\cite{Kendall1992} were used to accurately treat the van der Waals interactions. No Coulomb divergence corrections for the exchange and correlation self-energies were applied. The \gw@PBE valence band maximum (VBM) and conduction band minimum (CBM) energies of tested solids from the two schemes are presented in Table~\ref{tab:accd}.

\begin{table}[hbt]
	\centering
	\caption{\gw@PBE valence band maximum and conduction band minimum energies for several periodic crystals using a $4\times4\times4$ $\veck$-mesh (with no Coulomb divergence corrections). The ``VBM-AC'' and ``CBM-AC'' columns are QP energies from the  \gw-AC scheme, while the ``$\delta$VBM'' and ``$\delta$CBM'' columns refer to differences between the \gw-AC and \gw-CD QP energies ($\epsilon_\mathrm{AC}-\epsilon_\mathrm{CD}$). The unit of energy is eV.}
	\label{tab:accd}
	\begin{tabular}{>{\centering\arraybackslash}p{1.4cm}>{\centering\arraybackslash}p{1.7cm}>{\centering\arraybackslash}p{1.4cm}>{\centering\arraybackslash}p{1.7cm}>{\centering\arraybackslash}p{1.4cm}}
	\hline\hline
System	 &  VBM-AC  &  $\delta$VBM ($\times 10^{-3}$) & CBM-AC & $\delta$CBM ($\times 10^{-3}$)  \\
	\hline
Si   &   8.56 & 0  & 9.51 & 0   \\
C &  15.09 & 0 & 20.11 & -1   \\
SiC & 12.08 & -1 & 14.09 & -20  \\
AlP & 7.25 & 0 & 9.42 & -8 \\
BN & 12.65 & 0  & 18.27 & -37 \\
BP & 11.49 & -1 & 13.39 & -14\\
GaN & 12.23 & 0 & 15.13 & -14\\
GaP & 9.58 & 0 & 11.68 & -7\\
MgO & 8.71 & 0 & 15.33 & -11\\
ZnO & 9.08 & 0 & 11.78 & 0 \\
ZnS & 7.76 & 0 & 11.20 & -5 \\
Ne & -12.50 & 0 & 5.20 & 0 \\
Ar & -4.86 & 1 & 6.97 & 6\\
\hline
MAD & & \textbf{0} & & \textbf{10} \\
    \hline\hline
	\end{tabular}
\end{table}

We find that for the VBM energies, the \gw-AC results are in excellent agreement with the \gw-CD values. For the CBM energies, there is a mean absolute difference of only 0.01 eV, and the maximum difference is less than 0.04 eV, which is much smaller than the possible errors introduced from other factors (e.g., the finite basis and $\veck$-mesh). Therefore, we will use the more efficient \gw-AC approach for all studies of valence band energies in the subsequent sections.

\subsection{Basis Set Convergence}\label{sec:conv}

\begin{figure}[hbt]
\centering
\subfigure{
\includegraphics{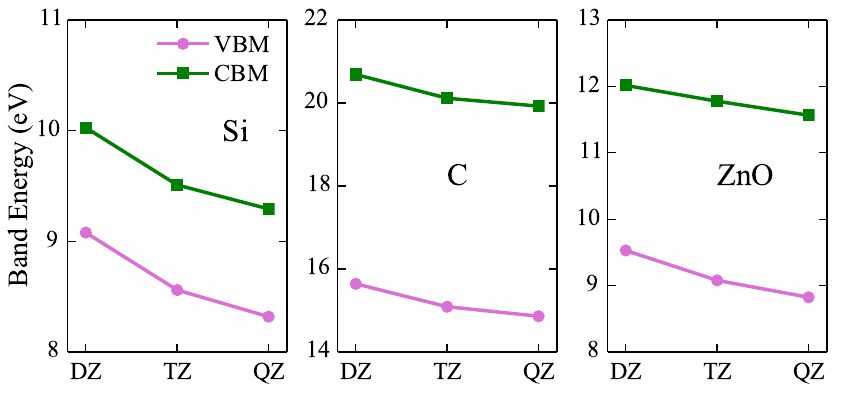}}
\vfill
\subfigure{
\includegraphics{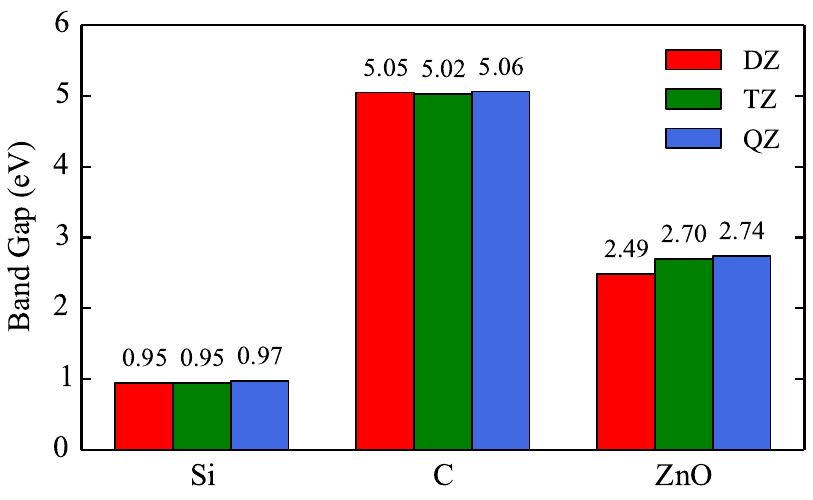}}
\caption{\gw@PBE valence band maximum and conduction band minimum energies (top) and band gaps (bottom) of Si, C and cubic ZnO using the cc-pVXZ (X=D, T, Q) basis sets and a $4\times4\times4$ $\veck$-mesh (no Coulomb divergence correction is applied). The band gap values are shown on top of the bars.}
\label{fig:basis}
\end{figure}

We now study how the \gw~band energies and band gaps of periodic systems converge with respect to the Gaussian basis set size.
This problem is especially interesting in transition metal containing systems because such systems normally require a large number of plane waves (and empty states) to converge the \gw~QP energies~\cite{Jiang2018,Rangel2020}. Here, we performed \gw@PBE calculations for the band gaps of Si, C and ZnO (zinc blende) using a $4\times4\times4$ $\veck$-mesh. The cc-pVXZ basis sets (X=D, T, Q) and their corresponding cc-pVXZ-RI auxiliary basis were used
because they are designed to capture correlation in a systematic fashion in wavefunction based quantum chemistry calculations.
The results are shown in Fig.~\ref{fig:basis}.

It can be seen that the \gw@PBE band energies (VBM and CBM) converge slowly as the basis size increases from DZ to QZ for all three solids tested. For Si, the VBM energy is 0.24 eV larger in the cc-pVTZ basis than in the cc-pVQZ basis. However, we also see that the VBM and CBM energies are changing in the same direction and have a surprisingly good error cancellation. As a result, the \gw@PBE band gaps of Si and C are already converged to within 0.04 eV of the largest basis result using the cc-pVDZ basis,
which only has 36 (Si) and 28 (C) Gaussian basis functions per unit cell. This behavior is very similar to the Gaussian-based \gw~results for molecules reported in Ref.~\cite{Wilhelm2016}, where the band gap of the benzene molecule is converged using an aug-DZVP basis, although the ionization potential is still far away from the basis set limit. On the other hand, the level of error cancellation may vary significantly based on the difference in excitation character between the valence and conduction bands in solids, as shown previously~\cite{VanSetten2017}. Therefore, it is interesting to investigate the Gaussian basis set convergence in a variety of solids with different excitation character in the future. 

For cubic ZnO, the cc-pVTZ basis is needed to reach an accurate band gap
compared to the cc-pVQZ value. However, we note that ZnO is a well-known challenging  system for \GW~calculations, largely because the \GW~self-energy converges very slowly with respect to the truncated number of virtual bands in the polarizability calculation~\cite{Shih2010,Friedrich2011,Stankovski2011,Rangel2020}. Typically, a few thousand plane-wave basis functions are required to avoid a substantial underestimation of the \gw~band gap for ZnO.
One approach to deal with this issue is to extend the LAPW method with high-energy local orbitals (HLOs), which reduces the number of plane-waves to a few hundred~\cite{Jiang2016}. In our Gaussian-based \gw~calculation for cubic ZnO, there are only 98 (cc-pVTZ) and 159 (cc-pVQZ) basis functions
per unit cell, which is much smaller than the basis size in purely plane-wave based calculations. This is because the correlation
consistent construction of the Gaussian bases
discretizes the virtual states in a way that is specifically designed to rapidly converge the correlation energy.
Due to this small basis size, no virtual band truncation is needed to compute the polarizability. This suggests that Gaussian bases
have an advantage as a compact and efficient choice when performing \gw~calculations on periodic systems. From the above analysis,
we will use the cc-pVTZ/cc-pVTZ-RI basis in the calculations that follow unless otherwise specified. 

\subsection{Finite Size Convergence}\label{sec:finite}
\begin{figure}[hbt]
\centering
\subfigure{
\includegraphics{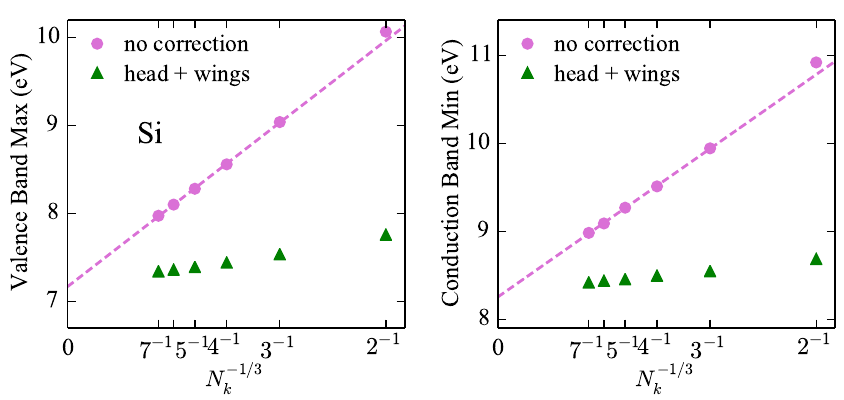}}
\vfill
\subfigure{
\includegraphics{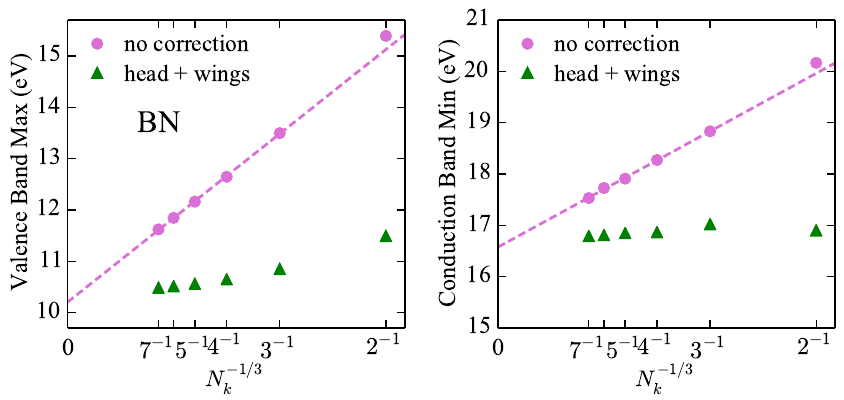}}
\vfill
\subfigure{
\includegraphics{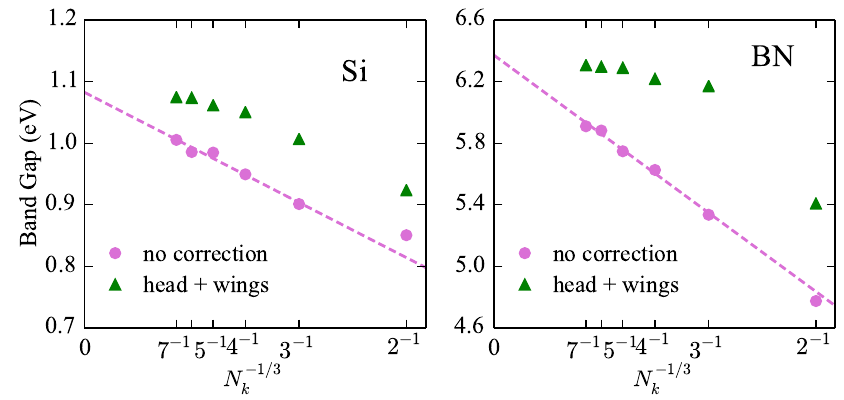}}%
\caption{\gw@PBE VBM, CBM and band gaps of Si and cubic BN as a function of $N_\veck^{-1/3}$ in the cc-pVTZ basis, with and without Coulomb divergence corrections. A linear extrapolation (dashed line) is performed on the uncorrected energies. Top: VBM and CBM of Si. Middle: VBM and CBM of cubic BN. Bottom: Band gaps of Si and BN.}
\label{fig:kpoint}
\end{figure}

We next investigate the convergence of \gw~valence excitation energies and band gaps with respect to the Brillouin zone sampling, i.e., the number of $\veck$-points $N_\veck$. As discussed in Section~\ref{sec:head}, without the $\mathbf{G} = \veczero$ contribution to the exchange and correlation \gw~self-energies, we expect the \gw~excitation energies and band gaps to have a finite-size error that scales like $\mathcal{O}(N_\veck^{-1/3})$. We thus plot the VBM, CBM and band gaps of Si and cubic BN computed using \gw@PBE in Fig.~\ref{fig:kpoint} as a function of $N_\veck^{-1/3}$, using the cc-pVTZ basis and increasing $\veck$-meshes from $2\times2\times2$ to $7\times7\times7$. As a comparison, we also show the \gw~energies obtained
by using the head and wings Coulomb divergence corrections to the correlation self-energy as well as the correction to the exchange self-energy.

From this data we see that the uncorrected \gw~VBM, CBM and band gaps of Si and cubic BN indeed depend approximately linearly on $N_\veck^{-1/3}$. We thus fit the uncorrected \gw~VBM, CBM and band gap values from the $3\times3\times3$ to $7\times7\times7$ $\veck$-meshes to the form $E(N_\veck^{-1/3}) = E_\infty + a N_\veck^{-1/3}$ to extrapolate to the thermodynamic limit (TDL). Compared to the extrapolated values, one can see that the finite size errors in
the VBM and CBM of Si and cubic BN are significantly reduced by applying the Coulomb divergence (head + wings) corrections. Even using the small $3\times3\times3$ $\veck$-mesh, the corrected VBM/CBM energies are very close to the TDL values.

The extrapolated \gw@PBE band gaps are 1.08 eV and 6.37 eV for Si and cubic BN respectively, in very good agreement with the experimental values (1.17 eV for Si~\cite{Madelung2004}, 6.4 eV for BN~\cite{Goldmann1989}). For BN, the Coulomb divergence corrected band gap converges to the TDL much faster than the uncorrected band gap. Using the $7\times7\times7$ $\veck$-mesh, the \gw@PBE band gap of cubic BN is 6.31 eV, which is only 0.06 eV smaller than the extrapolated TDL value. Similarly, for Si, the Coulomb divergence corrected band gaps also converge more quickly with the number of $\veck$-points compared to the uncorrected values. The corrected \gw~band gap of Si is 1.07 eV using the $7\times7\times7$ $\veck$-mesh, which is only 0.01 eV smaller than the extrapolated TDL value. Overall, we demonstrate that when using the computed Coulomb divergence corrections, one can obtain reasonably well-converged \gw~excitation energies with respect to the $\veck$-point sampling at an affordable cost.

\subsection{Benchmark of Band Gaps}\label{sec:bandgap}
In this section, we present our \gw-AC benchmark results for band gaps of 15 semiconductors and rare gas solids. In addition to the 13 solids described in Section~\ref{sec:accd}, we further include ZnO and AlN in the wurtzite structure, marked as wZnO and wAlN. For the 13 cubic semiconductors and rare gas solids, we computed the \gw@PBE band gaps using $\veck$-meshes ranging from $3\times3\times3$ to $6\times6\times6$ without Coulomb divergence corrections, then performed linear extrapolations of the form $E(N_\veck^{-1/3}) = E_\infty + a N_\veck^{-1/3}$ as described in Section~\ref{sec:conv} to obtain the band gaps in the TDL. For wZnO and wAlN, we used  $3\times3\times2$, $4\times4\times3$ and $6\times6\times4$ $\veck$-meshes, and then performed linear extrapolations for the \gw~band gaps. We also report the \gw@PBE band gaps obtained by applying Coulomb divergence corrections for the $6\times6\times6$ $\veck$-mesh ($6\times6\times4$ for wAlN and wZnO). cc-pVTZ/cc-pVTZ-RI basis sets were used for all solids, except
for Ne and Ar where we used the aug-cc-pVTZ/aug-cc-pVTZ-RI basis. All lattice constants and detailed band gap values for different $\veck$-meshes can be found in the Supporting Information.  The \gw@PBE band gaps are presented in Table~\ref{tab:bandgap}, and compared to DFT-PBE results and experimental values. In certain materials (e.g., zinc blende ZnO with Coulomb divergence correction), we found multiple solutions when solving the QP equation Eq.~\ref{eq:qp}. In such cases, we report the QP solution with the highest quasiparticle weight computed according to Eq.~\ref{eq:qpweight}.

\begin{table}[hbt]
	\centering
	\caption{Band gaps of semiconductors and rare gas solids from DFT-PBE ($6\times6\times6$ $\veck$-mesh), Coulomb divergence corrected \gw@PBE ($6\times6\times6$ $\veck$-mesh), $\veck$-point extrapolated \gw@PBE and experiments. The cc-pVTZ basis was used unless otherwise specified.
          MARE stands for mean absolute relative error compared to the experimental value. All band gap values are in eV.}
	\label{tab:bandgap}
	\begin{tabular}{>{\centering\arraybackslash}p{2cm}>{\centering\arraybackslash}p{1.5cm}>{\centering\arraybackslash}p{1.5cm}>{\centering\arraybackslash}p{1.6cm}>{\centering\arraybackslash}p{1.7cm}}
	\hline\hline
System	 &  PBE  & {\gw \newline ($6\times6\times6$)} & {\gw \newline (extrap.)} &  Expt.  \\
	\hline
Si   &   0.61 & 1.07 & 1.08  &  1.17~\cite{Madelung2004} \\
C &  4.14 & 5.55 & 5.52 &  5.48~\cite{Goldmann1989}  \\
SiC & 1.36 & 2.31  & 2.44 & 2.42 ~\cite{Goldmann1989}  \\
BN & 4.47  & 6.30 & 6.41  & 6.4 ~\cite{Goldmann1989} \\
BP & 1.42 & 2.10& 2.15 & 2.4 ~\cite{Goldmann1989} \\
wAlN~\textsuperscript{\emph{a}} & 4.19 & 5.85 & 5.89 & 6.2-6.3 ~\cite{Goldmann1989} \\
AlP & 1.62 & 2.34 & 2.41 & 2.51~\cite{Madelung2004} \\
GaN & 1.84 & 3.16 & 3.13 & 3.17 ~\cite{Madelung2004} \\
GaP & 1.69 & 2.24 & 2.33 & 2.27~\cite{Goldmann1989} \\
MgO~\textsuperscript{\emph{b}} & 4.75 & 7.46 & 7.43 & 7.83~\cite{Whited1973} \\
ZnO & 0.95 & 3.05 & 2.91 & 3.4~\textsuperscript{\emph{d}} \\
wZnO~\textsuperscript{\emph{a}} & 1.07 & 3.16 & 3.08 & 3.4~\cite{Goldmann1989} \\
ZnS & 2.36 & 3.69 & 3.63 & 3.7~\cite{Goldmann1989} \\
Ne~\textsuperscript{\emph{c}} & 11.65 & 19.77 & 20.01 & 21.7~\cite{Schwentner1975} \\
Ar~\textsuperscript{\emph{c}} & 8.77 & 13.09 & 13.24 & 14.2~\cite{Schwentner1975} \\
\hline
MARE (\%) & \textbf{42} & \textbf{5.5} & \textbf{5.2} &  \\
    \hline\hline
	\end{tabular}
	
\textsuperscript{\emph{a}} A $6\times6\times4$ $\veck$-mesh was used for the second (PBE) and third (\gw) columns.\\
\textsuperscript{\emph{b}} The most diffuse $p$ function of Mg was removed to avoid linear dependencies.\\
\textsuperscript{\emph{c}} aug-cc-pVTZ basis.\\
\textsuperscript{\emph{d}} The experimental band gap of zinc blende ZnO is not available, so we use the wurtzite ZnO gap as an approximation.
\end{table}

As shown in Table~\ref{tab:bandgap}, \gw@PBE improves the description of band gaps significantly over PBE, a finding which agrees with previous
studies. The mean absolute relative error (MARE) of \gw@PBE using the $\veck$-point extrapolation is only 5.2\% compared to the experimental values. The Coulomb divergence corrected \gw@PBE using the $6\times6\times6$ $\veck$-mesh achieves a comparable accuracy (MARE = 5.5\%). 

\begin{table}[hbt]
	\centering
	\caption{Band gaps of wZnO from PBE, \gw@PBE, LDA, and \gw@LDA calculations, with Coulomb divergence correction at $6\times6\times4$ $\veck$-mesh, using different Gaussian basis sets. All band gap values are in eV.}
	\label{tab:wzno}
	\begin{tabular}{>{\centering\arraybackslash}p{2.cm}>{\centering\arraybackslash}p{2cm}>{\centering\arraybackslash}p{2.2cm}>{\centering\arraybackslash}p{2.2cm}}
	\hline\hline
Method	 &  cc-pVTZ & def2-TZVPP &  cc-pVTZ-PP  \\
	\hline
PBE   &   1.07 & 1.07 & 0.92    \\
\gw@PBE &  3.16 & 3.16 & 2.85   \\
LDA & 0.78 & 0.78  & 0.58  \\
\gw@LDA & 3.06  & 3.02 & 2.78  \\
\hline\hline
	\end{tabular}
\end{table}

Special attention should be placed on wZnO, whose previously reported \gw~band gap values differ by more than 2 eV using different \gw~approximations and codes~\cite{Rangel2020}. A recent benchmark comparison of state-of-the-art full-frequency plane-wave \gw~codes reported that the \gw@LDA band gap of wZnO extrapolated to the basis set limit is 2.76 eV using a shifted $8\times8\times5$ $\veck$-mesh~\cite{Rangel2020}. The accurate LAPW+HLOs method in Ref.~\cite{Jiang2016} predicted the \gw@PBE band gap of wZnO to be 3.01 eV using a $6\times6\times4$ $\veck$-mesh. Our Gaussian-based \gw@PBE band gap of wZnO is 3.16 eV using the $6\times6\times4$ $\veck$-mesh, and the extrapolated TDL band gap is 3.08 eV, in good agreement with the LAPW+HLOs result. We have also computed the \gw@LDA band gap of wZnO (in a cc-pVTZ basis) to facilitate a direct comparison against Ref.~\cite{Rangel2020}, as shown in Table~\ref{tab:wzno}. Our \gw@LDA band gap of wZnO is 3.06 eV using the $6\times6\times4$ $\veck$-mesh and Coulomb divergence correction, and the extrapolated TDL result is 2.97 eV. These values are 0.2-0.3 eV larger than the 2.76 eV band gap reported in Ref.~\cite{Rangel2020}. 

We note that the PBE band gap of wZnO in the cc-pVTZ basis is 1.07 eV and larger than the 0.80-0.83 eV values computed using LAPW or PAW basis in Refs.~\cite{Jiang2016,Hinuma2014}. On the other hand, the LDA band gap of wZnO in the cc-pVTZ basis (0.78 eV) agrees well with reported values (0.73-0.75 eV)~\cite{Friedrich2011,Jiang2016}. To understand the difference in the PBE band gap, we performed calculations using two other Gaussian basis sets - def2-TZVPP~\cite{Weigend2005} (all-electron) and cc-pVTZ-PP~\cite{Peterson2005} (pseudopotential for $1s2s2p$ core electrons of Zn). As shown in Table~\ref{tab:wzno}, two all-electron Gaussian basis sets (cc-pVTZ and def2-TZVPP) produce almost the same band gaps for wZnO, while the cc-pVTZ-PP basis predicts 0.15/0.31 eV smaller PBE/\gw@PBE band gap. Thus, we speculate the difference between cc-pVTZ and LAPW/PAW bases on the PBE band gap is due to the different treatment of core electrons (for a more thorough comparison of Gaussian basis sets, see SI Table S2). Because difference at the PBE level can contribute to difference in the \gw~band gaps, we call for caution when comparing our \gw@PBE result against other \gw@PBE implementations on ZnO. Again, we emphasize that only 196 Gaussian basis functions
per wZnO unit cell were needed to obtain these results. Based on this benchmark, we conclude that our all-electron Gaussian-based \gw~implementation is
both accurate and efficient when performing \gw~calculations for valence excitations in periodic systems.

\subsection{\gw~for Metals}
Metallic systems require additional considerations in \gw~because of the vanishing energy gaps. At the mean-field level, finite temperature smearing is often applied to allow a Fermi-Dirac fractional occupation of the orbitals $\{f_{m\veck_m}\}$:
\begin{equation}\label{eq:fracocc}
f_{m\veck_m} = \frac{1}{1+e^{(\epsilon_{m\veck_m}-\mu)/\sigma}},
\end{equation}
where $\mu$ is the chemical potential and $\sigma$ is the finite temperature smearing parameter. In the \gw-AC scheme, Eq.~\ref{eq:Pi} becomes
\begin{equation}\label{eq:Pimetal}
\begin{split}
\vecPi_{PQ}(\vecq, i\omega') = \frac{1}{N_\veck} \sum_{i\veck_i a\veck_a}  
 v_{P}^{ia} \frac{(f_{i\veck_i} - f_{a\veck_a}) (\epsilon_{i\veck_i} - \epsilon_{a\veck_a})}{\omega'^2 + (\epsilon_{i\veck_i} - \epsilon_{a\veck_a})^2} v_{Q}^{ai} ,
\end{split}
\end{equation}
and the sum over states $i\veck_i$ and $a\veck_a$ runs over all molecular orbitals as all orbitals are partially occupied. This means an electron may be excited within the same energy band, which is termed an intraband transition. $\epsilon_F$ in Eq.~\ref{eq:sigmaAC3} is set to $\mu$. For metallic systems, intraband transitions lead to a non-vanishing Drude term in the long-wavelength limit, which in principle may be included as a type of
finite size correction~\cite{Liu2016}. However, we have not implemented this correction term here and leave it to future work.

\begin{figure}[hbt]
\centering
\includegraphics{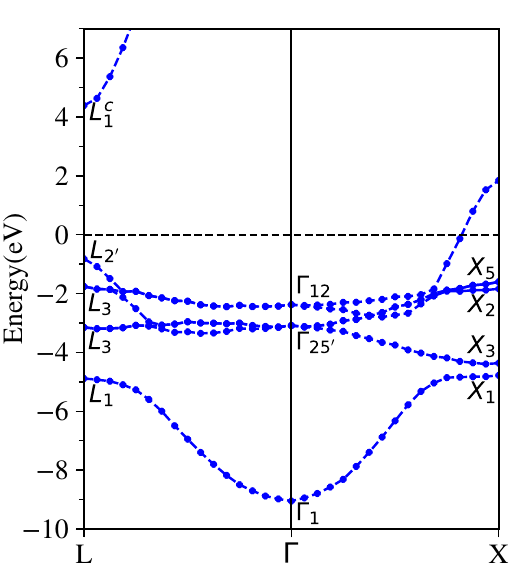}
\caption{Band structure of bulk Cu computed by Gaussian-based \gw@PBE using the def2-TZVP basis and an $8\times8\times8$ $\veck$-mesh.}
\label{fig:cu}
\end{figure}

We applied our all-electron Gaussian-based \gw~method to bulk Cu in the fcc structure (lattice constant 3.603~\AA~\cite{Schimka2011}). The band structure of Cu was computed using PBE and \gw@PBE with the def2-TZVP/def2-TZVP-RIFIT basis~\cite{Weigend2005,Hattig2005} and an $8\times8\times8$ $\veck$-mesh. No finite size corrections were applied here since we have not yet implemented the intraband transition Drude term. A Fermi-Dirac finite temperature smearing ($\sigma=0.002$ a.u.) was used in the PBE calculation. We note that for metallic systems, one may need more modified Gauss-Legendre grids for stable numerical integration of Eq.~\ref{eq:sigmaAC3}. For Cu, we used 200 grid points. The results are shown in Fig.~\ref{fig:cu} and Table~\ref{tab:cu} and compared to previous \gw~studies and experimental values.

\begin{table*}[hbt]
	\centering
	\caption{Band energies and bandwidths of bulk Cu calculated by PBE and all-electron Gaussian-based \gw@PBE using the def2-TZVP basis and $8\times8\times8$ $\veck$-mesh (with no finite size corrections). Special symmetry points are noted in Fig.~\ref{fig:cu}. The results are compared to \gw@PBE results using the PAW method in Ref.~\cite{Liu2016} and \gw@LDA results using the pseudopotential plane-wave (PPW) scheme in Ref.~\cite{Marini2002}, as well as the experimental values~\cite{Courths1984}.}
	\label{tab:cu}
	\begin{tabular}{>{\centering\arraybackslash}p{3.8cm}>{\centering\arraybackslash}p{1.6cm}>{\centering\arraybackslash}p{1.4cm}>{\centering\arraybackslash}p{2.0cm}>{\centering\arraybackslash}p{1.5cm}>{\centering\arraybackslash}p{1.5cm}>{\centering\arraybackslash}p{1.5cm}}
	\hline\hline
	&  &  PBE  & \gw@PBE & PAW~\cite{Liu2016} & PPW~\cite{Marini2002} & Expt.~\cite{Courths1984} \\
	\hline
Positions of $d$ bands &   $\Gamma_{12}$ & -2.18 & -2.38 & -2.11 & -2.81 & -2.78  \\
 & $X_5$ & -1.46 & -1.60 & -1.45 & -2.04  & -2.01 \\
 & $L_3$ & -1.60 & -1.76 & -1.58 & -2.24  & -2.25 \\ 
 \hline
Widths of $d$ bands &   $\Gamma_{12} - \Gamma_{25'}$ & 0.83 & 0.71 & 0.69 & 0.60  & 0.81  \\
 &   $X_{5} - X_{3}$ & 2.97 &  2.76  & 2.60 & 2.49 & 2.79 \\
 &   $X_5 - X_1$ & 3.41 &  3.18 & 3.10 & 2.90 & 3.17 \\
 &   $L_3 - L_3 $ & 1.44 &  1.39  & 1.26 & 1.26  & 1.37\\
  &   $L_3 - L_1$ & 3.42 &  3.13  & 3.16 & 2.83  & 2.91\\ 
  \hline
Positions of $s/p$ bands &   $\Gamma_1$ & -9.14 &  -9.05 & -9.18 & -9.24  & -8.60 \\
&   $L_{2'}$ & -0.80 & -0.82 & -1.02 & -0.57  & -0.85 \\
\hline
$L$ gap &   $L_{1}^c - L_{2'}$ & 4.89 & 5.21 & 4.98 & 4.76 & 4.95 \\
    \hline\hline
	\end{tabular}
\end{table*}

As shown in Table~\ref{tab:cu}, PBE predicts larger $d$ bandwidths than the experimental values. For example, the $X_5 - X_1$ and $L_3 - L_1$ bandwidths are 0.24 eV and 0.51 eV too wide. Our Gaussian-based \gw@PBE approach narrows the $d$ bandwidths by 0.1-0.3 eV, leading to better agreement with the experimental values than PBE. Our \gw@PBE $d$ bandwidths are also closer to the experimental values (except for $L_3 - L_1$) compared to plane-wave based \gw~results in PAW~\cite{Liu2016} and pseudopotential PW (PPW)~\cite{Marini2002} schemes. The PBE positions of the $d$ bands and $\Gamma_1$ are predicted to be 0.55-0.65 eV too shallow and 0.54 eV too deep compared to experiment. Our \gw@PBE corrects these positions by ~0.1-0.2 eV, but this is still far from quantitative agreement
with experiment. This is similar to the results obtained in the PAW scheme.
The much better performance of PPW-based \gw@LDA for the positions of the $d$ bands should be attributed to the fortuitous error cancellation between the applied pseudopotential and the \gw~approximation, as discussed in Ref.~\cite{Liu2016}. Finally, our \gw@PBE $L$ gap is 0.26 eV larger than the experimental value, and worse than that obtained from PBE and other \gw~schemes. We believe this is largely due to the missing finite size
corrections in our calculation. 

\subsection{Computational Efficiency}
\begin{figure}[hbt]
\centering
\includegraphics{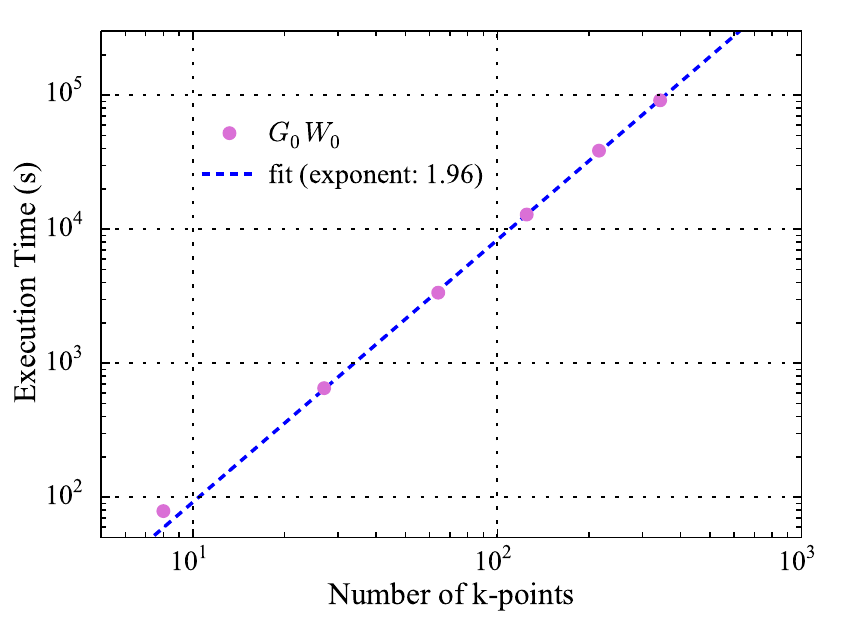}
\caption{Execution (wall) time of \gw@PBE calculations of Si in the cc-pVTZ basis, as a function of the number of $\veck$-points, on a 28-core CPU node. Six quasiparticle energy levels are computed for each $\veck$-point in each case.}
\label{fig:time}
\end{figure}

We further report on the computational efficiency of our Gaussian-based \gw~Python code. In Fig.~\ref{fig:time}, we present the execution (wall) time of \gw@PBE calculations of Si in the cc-pVTZ basis on a 28-core CPU node, with threading handled by OpenMP. As can be seen, the $6\times6\times6$ Si calculation takes about 11 hours to complete on a single node. Consistent with the discussion in Section~\ref{sec:AC}, the computational cost has approximately an $\mathcal{O}(N_\veck^2)$ scaling with respect to the number of $\veck$-points. We note that another prominent computational step in terms of cost is the evaluation of GDF integrals prior to the $\gw$~calculation, which takes 7 hours on a single node for the $6\times6\times6$ Si case using current PySCF implementation. We store the precomputed GDF integrals on disk, which requires 390 GB disk space. The memory requirement is much lower,  scaling as $\mathcal{O}(N_\veck N_{AO}^2 N_{aux})$; this amounts to 11 GB for the same calculation.

\section{Core Excitation Binding Energies}
Core-level X-ray photoelectron spectroscopy (XPS) is a powerful tool for chemical analysis in solids and surfaces~\cite{Fadley2010}, because core
excitations are sensitive to the atomic environment. A key challenge when using XPS in complex materials is that the assignment of the
experimental XPS peaks to the specific atomic sites is often difficult. Thus, it is of great interest to develop accurate first-principles
approaches to simulate core excitation binding energies (CEBEs). 

One state-of-the-art method for calculating CEBEs is the $\Delta$-self-consistent-field ($\Delta$SCF) approach based on KS-DFT~\cite{Besley2009}, which has been shown to predict accurate CEBEs with approximately 0.2-0.3 eV errors for small molecules~\cite{PueyoBellafont2016,Kahk2019}. Recently, a restricted open-shell Kohn-Sham (ROKS) approach~\cite{Hait2020} has also been proposed which achieves similar accuracy in molecules. However, $\Delta$SCF methods
are more complicated to apply to periodic systems, due to the need to treat the Coulomb divergence when one electron is explicitly removed from the system. In practice, some ways to proceed include using finite cluster models~\cite{Kahk2019}, adding an excess electron into the conduction bands~\cite{Pehlke1993} or using an exact Coulomb cutoff method~\cite{Ozaki2017}.

\GW~methods provide an alternative treatment of this problem. By computing the core-state quasiparticle energies through \GW, one naturally incorporates
relaxation and correlation effects into the core excitation energies~\cite{VanSetten2018}. Because \GW~does not explicitly create a core hole, no special treatment of periodic boundary conditions needs to be used for solids unlike in $\Delta$SCF methods. Recently, Golze \textit{et al.}~\cite{Golze2018a,Golze2020a} demonstrated that \GW~in a Gaussian basis predicts accurate absolute core-level binding energies in XPS with approximately 0.3 eV errors in
molecules, when using an eigenvalue self-consistent \textit{G}\textit{W}\textsubscript{0} approach or \gw~combined with a density functional with a large percentage of exact exchange. 
\GW~in combination with the Bethe-Salpeter equation (\GW+BSE) has also been widely used for simulating X-ray absorption (XAS) and X-ray emission (XES) core excitation spectra in molecules and solids~\cite{Vinson2011,Gilmore2015}.
On the other hand, there are relatively few studies that have applied the \GW~approach to investigate the XPS core excitation energies in solids~\cite{aoki2018,ishii2010}. Aoki \textit{et al.}~\cite{aoki2018} showed that \gw~starting from a self-interaction-corrected LDA functional predicted XPS CEBEs within 1 eV of error for a few semiconductors, using a mixed plane-wave and numerical atom-centered orbital basis scheme.
Our all-electron Gaussian-based \gw~code is well placed to model core-electron physics, so we apply this method to compute XPS CEBEs in semiconductor materials with explicit periodic boundary conditions.

As discussed in Ref.~\cite{Golze2018a}, the frequency structure of the core-state \gw~self-energy is very complicated and cannot be accurately reproduced by a simple Pad\'{e} analytic continuation. Thus, in this section, we choose to use the \gw-CD approach that directly works on the real axis, and
always solve the QP equations self-consistently. One more complexity in gapped solids comes from the uncertainty of the Fermi level in experiments. Unlike in molecules, the Fermi level may be anywhere in the gapped region in solids, and may be pinned by possible defects in the system, leading to large variations in the experimental core binding energies~\cite{Walter2019}. For instance, the reported experimental C $1s$ core binding energy in diamond varies from 283.25 eV to 291.35 eV~\cite{Walter2019}, making  benchmarking of theoretical methods impossible. One way to deal with this issue is to redefine the core excitation binding energy (CEBE) as the difference between the core binding energy and the valence band maximum energy of solids~\cite{aoki2018}:
\begin{equation}\label{eq:cebe}
\mathrm{CEBE} = | \epsilon^{\mathrm{core}} - \epsilon^{\mathrm{VBM}} |.
\end{equation}
In this way, the ambiguity in the experimental Fermi level is removed. We thus use this definition of CEBEs to benchmark our \gw~implementation.

We computed 14 CEBEs (the $1s$ level of C, N, O, Be, B and $2p$ level of Al, Si, Mg) in typical semiconductors using DFT-PBE and finite-size-corrected \gw@PBE methods. We also tested the performance of a hybrid functional PBE45 (where 45\% of HF exchange is used) and \gw@PBE45. This choice was suggested in Ref.~\cite{Golze2020a}, where the authors found \gw@PBE45 gave the best \gw~results for core binding energies in molecules. To accurately describe the core states, we used the cc-pCVTZ basis set~\cite{Woon1995,Peterson2002} (which adds extra tight core basis functions to cc-pVTZ) for all elements, except for Zn and Ga where we used the cc-pVTZ basis. The cc-pwCVTZ-RI~\cite{Hattig2005}/cc-pVTZ-RI basis sets were used as auxiliary bases. $4\times4\times4$, $4\times4\times3$ and $4\times4\times2$ $\veck$-meshes were used for cubic, wurtzite and hexagonal materials respectively.  Relativistic effects were neglected in this study.

\begin{table*}[hbt]
	\centering
	\caption{Core excitation binding energies (defined in Eq.~\ref{eq:cebe}) calculated by PBE, PBE45 (HF exchange = 45\%), all-electron Gaussian-based \gw@PBE and \gw@PBE45 (with Coulomb divergence corrections). The cc-pCVTZ basis set was used for all elements, expect  for Zn and Ga, where the cc-pVTZ basis was used. $4\times4\times4$, $4\times4\times3$ and $4\times4\times2$ $\veck$-meshes were used for cubic, wurtzite and hexagonal materials respectively. Diamond, SiC, MgO, AlP and Si are in the cubic structure. AlN, GaN, BeO and ZnO are in the wurtzite structure. BN is in the hexagonal structure. MAE stands for  mean absolute error. All energies are in eV.}
	\label{tab:cebe}
	\begin{tabular}{>{\centering\arraybackslash}p{1.4cm}>{\centering\arraybackslash}p{1.6cm}>{\centering\arraybackslash}p{1.6cm}>{\centering\arraybackslash}p{2.1cm}>{\centering\arraybackslash}p{1.6cm}>{\centering\arraybackslash}p{2.1cm}>{\centering\arraybackslash}p{3cm}}
	\hline\hline
 Core	& Material &  PBE  & \gw@PBE & PBE45 & \gw@PBE45 & Expt. \\
	\hline
C $1s$ &   Diamond  & 265.66 & 276.49 & 280.41 & 284.59 & 283.7, 283.9~\cite{Goldmann1989}  \\
     &   SiC & 263.36 & 272.71 & 277.98 & 281.67 & 281.45~\cite{Waldrop1990}  \\ 
     \hline
N $1s$ &   BN  & 374.20 & 387.02 & 391.45 & 395.44 & 396.1~\cite{Hamrin1970}  \\ 
 &  AlN  &  373.44 &	383.14 &	390.64 &	394.38 &	393.87~\cite{Veal2008}  \\ 
 &  GaN  &  374.75 &	384.08 &	392.05 &	395.57 &	395.2~\cite{Duan2013} \\ 
 \hline
O $1s$ &  BeO  &  502.92	 & 514.63 &	522.38 &	526.84 &	527.7~\cite{Hamrin1970}  \\
 &  MgO  &  502.76 &	513.00 &	522.26 &	526.38 &	527.28~\cite{Chellappan2013}  \\
 &  ZnO   &  504.31 &	514.82 &	523.52 &	527.27 &	527.45~\cite{Veal2008}  \\
 \hline
Be $1s$ &  BeO  &  96.66 &	104.54 &	104.17 &	108.84 &	109.8~\cite{Hamrin1970}  \\
B $1s$ &  BN   &  171.69 &	183.33 &	183.18 &	187.72 &	188.4~\cite{Hamrin1970}  \\ 
\hline
Al $2p$ &  AlP   & 64.00 &	68.70 &	71.09 &	72.60 &	72.43~\cite{Waldrop1993} \\
 &  AlN  &  62.18 &	67.41 &	68.48 &	70.56 &	70.56~\cite{Veal2008}  \\ 
\hline
Si $2p$ &  Si  &  89.84 &	95.01 &	98.26 &	99.60 &	98.95~\cite{Yu1990}  \\
Mg $2p$ &  MgO   & 39.13 &	43.97 &	43.63 &	45.71 &	46.71~\cite{Liu2016a}, 46.79~\cite{Craft2007}   \\ 
\hline
MAE & & \textbf{16.77} & \textbf{7.92} & \textbf{3.59} & \textbf{0.57} \\
     \hline\hline
	\end{tabular}
\end{table*}

As presented in Table~\ref{tab:cebe}, CEBEs computed using PBE orbital energies are systematically smaller than the experimental values by over 16 eV. This large error arises because the KS orbital energies miss the orbital relaxation effects in the final core ionized state. \gw@PBE reduces the errors
of PBE significantly, but still has a mean absolute error (MAE) of 7.92 eV. The errors are even larger if only $1s$ CEBEs are considered. This unsatisfactory performance of \gw@PBE was also observed in molecules~\cite{VanSetten2018}, where the authors found linearized \gw@PBE gave errors as large as 9 eV in
the  core binding energies of molecules.  Switching to PBE45, the CEBE results are greatly improved over PBE and are even better
than \gw@PBE, which can be attributed to better error cancellation between the KS core orbital energies and neglecting the
orbital relaxation effects after core ionization. Using PBE45 as the starting point, \gw@PBE45 further reduces the MAE to only 0.57 eV, and the performance is equally good for the $1s$ (in C, N, O, Be, B) and $2p$ (in Al, Si, Mg) CEBEs. We thus conclude that when combined with PBE45,
Gaussian-based \gw~shows promise for the accurate simulation of CEBEs in periodic systems.

However, we also notice that the current \gw@PBE45 method does not always predict the correct relative CEBEs of the same core orbital in different materials. For example, \gw@PBE45 predicts the O $1s$ CEBE in BeO to be 0.43 eV smaller than in ZnO, while in experiments the O $1s$ CEBE in BeO is 0.25 eV larger. Such a discrepancy might be partly due to uncertainties in extracting the accurate valence band maximum in XPS experiments~\cite{Chambers2004}.
On the computational side, a key candidate for the source of error is the 
dependence of \gw~on the quality of the starting density functional. As shown in Table~\ref{tab:cebe}, PBE45 predicts the relative O $1s$ CEBE shift to be $-1.14$ eV (CEBE(BeO) $-$ CEBE(ZnO)), compared to $+0.25$ eV in experiments. Although \gw@PBE45 corrects the PBE45 CEBE shift to $-0.43$ eV, the result is still of the wrong sign. Therefore, it is interesting to study whether self-consistent \GW~(e.g., eigenvalue self-consistent {{\textit{G}\textit{W}\textsubscript{0}}}~\cite{Golze2020a} and quasiparticle self-consistent \GW~\cite{VanSetten2018})
can produce better relative CEBEs for solids in future work.

\section{Conclusions}
In this work, we described an all-electron \gw~implementation based on crystalline Gaussian basis sets for periodic systems and
benchmarked it on a range of systems including molecules, semiconductors, rare gas solids and a metal, for both valence and core excitations.
We demonstrated that modern Gaussian bases are a useful choice for carrying out periodic \gw~calculations,
finding that  \gw~band gaps
are rapidly converged using a small number of basis functions, as seen in the challenging case of ZnO. We developed a finite size correction
scheme similar to that used in plane-wave \gw~implementations, allowing our Gaussian-based \gw~calculations to converge to the thermodynamic limit
using a moderate amount of $\veck$-point sampling. We also investigated the performance of the \gw~approximation for core excitation binding
energies in semiconductors,
obtaining promising results in combination with a hybrid density functional with a large fraction of HF exchange. We conclude that the  Gaussian-based \gw~approach is a competitive choice for computing both valence and core excitation energies
in weakly-correlated materials. In addition, we want to point out that currently, there exists no Gaussian basis set optimized for correlated methods in periodic systems. It is of high interest to develop such Gaussian basis sets, considering that many correlated methods are now being developed
  for solid-state calculations. Our future work will examine the extension of the current scheme to
different types of self-consistency in \GW, as well as the combination of \gw~with other quantum chemistry methods, for example
using our recent full cell quantum embedding framework~\cite{Zhu2020a}.

\begin{acknowledgement}
 This work was supported by the US Department of Energy, Office of Science under award no. 19390.
TZ thanks Zhihao Cui, Yang Gao and Timothy Berkelbach for helpful discussions. Additional support was provided by the Simons Foundation via the Simons Collaboration on the Many Electron Problem, and via the Simons Investigatorship in Physics.
\end{acknowledgement}

\begin{suppinfo}
\gw@PBE band gaps of semiconductors and rare gas solids at various $\veck$-meshes used for finite size extrapolation. LDA and PBE band gaps of zinc blende ZnO using different Gaussian basis sets.
\end{suppinfo}

\providecommand{\latin}[1]{#1}
\makeatletter
\providecommand{\doi}
  {\begingroup\let\do\@makeother\dospecials
  \catcode`\{=1 \catcode`\}=2 \doi@aux}
\providecommand{\doi@aux}[1]{\endgroup\texttt{#1}}
\makeatother
\providecommand*\mcitethebibliography{\thebibliography}
\csname @ifundefined\endcsname{endmcitethebibliography}
  {\let\endmcitethebibliography\endthebibliography}{}

\end{document}